\documentclass[useAMS,usenatbib,aas_macros]{mn2e}
\usepackage{amsmath}
\usepackage{graphicx} 

\usepackage{array}
\usepackage{multirow}

\PassOptionsToPackage{normalem}{ulem}
\usepackage{ulem}

\usepackage{color}

\newcommand{\equ}[1]{eq.~(\ref{eq:#1})}

\newcommand{\Equ}[1]{Eq.~(\ref{eq:#1})}

\newcommand{\se}[1]{\S\ref{sec:#1}}
\newcommand{\fig}[1]{Fig.~\ref{fig:#1}}
\newcommand{\figs}[2]{Figs.~\ref{fig:#1}~-~\ref{fig:#2}}
\newcommand{\be}{\begin{equation}}
\newcommand{\ee}{\end{equation}}
\newcommand{\ba}{\begin{align}}
\newcommand{\ea}{\end{align}}
\newcommand{\bad}{\begin{equation} \begin{aligned}}
\newcommand{\ead}{\end{aligned} \end{equation}}
\newcommand{\bea}{\begin{eqnarray}}
\newcommand{\eea}{\end{eqnarray}}

\newcommand{\Msun}{M_\odot}

\newcommand{\ifm}[1]{\relax\ifmmode#1\else$\mathsurround=0pt #1$\fi}
\newcommand{\kms}{\ifmmode\,{\rm km}\,{\rm s}^{-1}\else km$\,$s$^{-1}$\fi}

\newcommand{\ltsima}{$\; \buildrel < \over \sim \;$}
\newcommand{\lsim}{\lower.5ex\hbox{\ltsima}}
\newcommand{\gtsima}{$\; \buildrel > \over \sim \;$}
\newcommand{\gsim}{\lower.5ex\hbox{\gtsima}}

\def\Msun{M_\odot}
\def\Msunh{h^{-1} M_\odot}

\def\omm{\Omega_{\rm m}}
\def\oml{\Omega_{\Lambda}}
\def\omb{\Omega_{\rm b}}

\def\Mv{M_{\rm vir}}
\def\Mvir{M_{\rm vir}}

\def\Rv{R_{\rm vir}}
\def\Vv{V_{\rm vir}}

\def\Ms{M_\star}
\def\Mstar{M_\star}

\def\Re{R_{\rm e}}
\def\Rd{R_{\rm d}}

\def\Rs{r_{\rm s}}
\def\tv{t_{\rm vir}}

\def\Rd{R_{\rm d}}

\def\Vc{V_{\rm c}}
\def\Vrot{V_{\rm rot}}

\def\lamh{\lambda_{\rm halo}}
\def\lamdm{\lambda_{\rm dm}}
\def\lamgas{\lambda_{\rm gas}}
\def\lams{\lambda_{\rm star}}

\def\lamgal{\lambda_{\rm gal}}
\def\jjh{\mathbf{j}_{\rm halo}} 
\def\jjgal{\mathbf{j}_{\rm gal}} 
\def\jjgas{\mathbf{j}_{\rm gas}} 
\def\jjstar{\mathbf{j}_{\rm star}} 
\def\jh{j_{\rm halo}} 
\def\jgal{j_{\rm gal}} 
\def\jgas{j_{\rm gas}} 
 
\def\jd{j_{\rm d}} 
\def\fj{f_j}

\def\tbn{t_{\rm BN}}
\def\thm{t_{\rm HM}}

\def\rms{{\rm RMS }}

\newcommand{\apjs}{ApJS}
\newcommand{\apjl}{ApJL}

\newcommand{\aap}{A\&A}

\title[Spins of Galaxies and Haloes]
{ Is the dark-matter halo spin a predictor of galaxy spin and size?}

\author[F. Jiang et al.]
{\parbox[t]{\textwidth}
{
Fangzhou Jiang$^{1}$\thanks{E-mail: fangzhou.jiang@huji.ac.il},
Avishai Dekel$^{1,2}$\thanks{E-mail: dekel@huji.ac.il},
Omer Kneller$^{1}$,
Sharon Lapiner$^{1}$,
Daniel Ceverino$^{3}$,
Joel R. Primack$^{4}$,
Sandra M. Faber$^{5}$,
Andrea Macci\`o$^{6,7}$,
Aaron Dutton$^{6}$,
Shy Genel$^{8,9}$,
Rachel S. Somerville$^{8,10}$
}
\\ \\ 
$^1$Center for Astrophysics and Planetary Science, Racah Institute of Physics,
The Hebrew University, Jerusalem 91904, Israel\\
$^2$SCIPP, University of California, Santa Cruz, CA 95064,
USA\\
$^3$Universit${\ddot{a}}$t Heidelberg, Zentrum f${\ddot{u}}$r Astronomie,
Institut f${\ddot{u}}$r Theoretische Astrophysik, Albert-Ueberle-Str. 2, 69120
Heidelberg, Germany\\
$^4$Department of Physics, University of California, Santa Cruz, CA 95064,
USA\\
$^5$Department of Astronomy, University of California, Santa Cruz, CA 95064,
USA\\
$^6$New York University Abu Dhabi, PO Box 129188, Abu Dhabi, United Arab Emirates\\
$^7$Max Planck Institute f${\ddot{u}}$r Astronomie, K${\ddot{o}}$nigstuhl 17, D-69117 Heidelberg, Germany\\
$^8$ Center for Computational Astrophysics, Flatiron Institute, 162 Fifth Avenue, New York, NY 10010, USA\\
$^9$ Columbia Astrophysics Laboratory, Columbia University, 550 West 120th Street, New York, NY 10027, USA\\
$^{10}$ Department of Physics and Astronomy, Rutgers, The State University of New Jersey, 136 Frelinghuysen Rd, Piscataway, NJ 08854, USA\\
}

\begin{document}

\large

\pagerange{\pageref{firstpage}--\pageref{lastpage}} \pubyear{2017}

\maketitle

\label{firstpage}

\begin{abstract}
The similarity between the distributions of spins for galaxies ($\lamgal$) and for dark-matter haloes ($\lamh$), indicated both by simulations and observations, is naively interpreted as a one-to-one correlation between the spins of a galaxy and its host halo. 
This is used to predict galaxy sizes in semi-analytic models via $\Re \simeq \lamh \Rv$, where $\Re$ is the half-mass radius of the galaxy and $\Rv$ is the halo radius.
Utilizing two different suites of zoom-in cosmological simulations, we find that $\lamgal$ and the $\lamh$ of its host halo are in fact only barely correlated, especially at $z \geq 1$. 
A general smearing of this correlation is expected based on the different spin histories, where the more recently accreted baryons through streams gain and then lose significant angular momentum compared to the gradually accumulated dark matter. 
Expecting the spins of baryons and dark matter to be correlated at accretion into $\Rv$, the null correlation at the end reflects an anti-correlation between $\lamgal/\lamh$ and $\lamh$, which can partly arise from mergers and a compact star-forming phase that many galaxies undergo.
On the other hand, the halo and galaxy spin vectors tend to be aligned, with a median $\cos \theta = 0.6$-0.7 between galaxy and halo, consistent with instreaming within a preferred plane.
The galaxy spin is better correlated with the spin of the inner halo, but this largely reflects the effect of the baryons on the halo.
Following the null spin correlation, $\lamh$ is not a useful proxy for $\Re$.
While our simulations reproduce a general relation of the sort $\Re = A \Rv$, in agreement with observational estimates, the relation becomes tighter with 
$A = 0.02 (c/10)^{-0.7}$, 
where $c$ is the halo concentration, which in turn introduces a dependence on mass and redshift. 
\end{abstract}

\begin{keywords}
{dark matter ---
galaxies: evolution ---
galaxies: formation ---
galaxies: haloes}
\end{keywords}

\section{Introduction}
\label{sec:intro}

\smallskip
In the standard paradigm of hierarchical structure formation, the angular momentum (AM) growth of a dark matter protohalo is driven by the large-scale gravitational tidal torque until maximum expansion. 
This ``tidal torque theory (TTT)'' (e.g., \citealt{doroshkevich70}; \citealt{white84}) has been shown to agree well with the predictions from cosmological $N$-body simulations (e.g., \citealt{porciani02a,porciani02b}).
The AM of a halo is often characterized using the dimensionless spin parameter \citep{bullock01j}: 
\be \label{eq:spin}
\lambda = \frac{j}{\sqrt{2}\Vv\Rv}
\ee
where $j=J/M$ is the specific angular momentum (sAM) and where $\Rv$ and $\Vv$ are the virial radius and velocity of the halo.
\footnote{\cite{peebles69} originally defined the spin parameter for dark matter haloes as 
$\lambda = J|E|^{1/2}G^{-1}M^{-5/2}$,
where $J$ is the magnitude of the total dark-matter AM within the virial radius; $M$ is the virial mass; and $E$ is the total energy of the halo.
Calculating the total energy $E$ is computationally expensive, and the potential energy is not uniquely defined, hindering practical use of the definition.
For smooth spherical haloes with all particles on circular orbits, the two definitions are linked via $\lambda_{\rm Peebles} = \lambda_{\rm Bullock} f_c^{-1/2}$, with $f_c$ measuring the deviation of $E$ from that of a singular isothermal sphere. 
The \citeauthor{bullock01j} definition enables one to compare the sAM of different components in a halo; while the \citeauthor{peebles69} definition can only be used considering all the mass inside a halo.}
The spin parameter as defined in \equ{spin} can refer to any part of the halo, and to any component of the galaxy inside the halo, such as the gas and/or stars in the disc and/or the whole galaxy, using the specific angular momentum of the component of interest. 

\smallskip
Galaxies form as cold gas condenses within the potential wells of dark matter haloes \citep{wr78,blum84}. 
Since the baryons and the dark matter have similar spatial distributions in the cosmic web, they are expected to gain comparable amount of sAM through the large-scale tidal torques. 
It used to be commonly assumed that the AM of the gas is conserved during the collapse, such that the galactic disc is expected to have a similar spin to that of its hosting halo $\lamgal\simeq\lamh$ (\citealt{fe80}; \citealt{mmw98}; \citealt{bullock01j}). 
In addition, the rotation curves of galaxies are close to flat, so for disc galaxies the rotation speed at some characteristic radius of the galaxy, $\Vrot$, is comparable to the virial velocity $\Vv$. 
These establish a link between the characteristic radius of the galaxy and the the virial radius of the halo.
The sAM of a galaxy can be written as
\be \label{eq:jgal}
\jgal \simeq \Re\Vrot ,
\ee
where $\Re$ is the 3D half stellar mass radius of a galaxy.\footnote{The more familiar form of \equ{jgal} in the literature (e.g., \citealt{mmw98}) is, for exponential discs, 
\be
\jd = 2\Rd\Vc, \nonumber
\ee
where $\jd$ is the sAM of the disc, $\Rd$ is the scale radius, and the rotation curve is flat at the level $\Vc$. 
Here we opt for a more general expression.
For exponential discs, $\Re\approx1.68\Rd$.}
The radius of the galaxy can be expressed as 
\begin{align} 
\Re & \simeq \frac{\jgal}{\jh} \frac{\jh}{\Rv\Vv}\frac{\Vv}{\Vrot}\Rv \nonumber \\ 
& = \fj \lamh \Rv. \label{eq:Re1}
\end{align}
Here the factor $\fj\equiv \lamgal/\lamh$, with $\lamgal$ denoting $\jgal/(\sqrt{2}\Rv\Vv)$, is taken as unity if the gas sAM is assumed to be conserved.
This is adopted in most semi-analytic models of galaxy formation (e.g., \citealt{somerville08b}; \citealt{guo11}; \citealt{benson12}) when trying to predict disc sizes.

\smallskip
Cosmological $N$-body simulations show that halo spin follows log-normal distributions with a median of $\langle\lamh\rangle\simeq0.035$ and a standard deviation of $\sigma_{\log_{10}\lamh} \simeq 0.25$ (e.g., \citealt{bullock01j}, \citealt{bett07}).
Hydro-cosmological simulations show that $\lamgal$ and $\lamh$ follow similar, log-normal distributions\footnote{though there are differences between the distributions of spins of the different components of the galaxy}.
For example, the spin of gas within $0.1\Rv$ has a median value of $\sim0.04$ and a standard deviation of $\sim$0.25 dex (e.g., \citealt{danovich15}).
Observed star-forming disc galaxies in the redshift range $z=0.5$-3 indicate a similar result \citep{burkert16}. 
The spin parameters of the discs were evaluated using their sAM based on measured H$_\alpha$ kinematics and normalized by the halo virial velocities and radii inferred from the kinematics and stellar mass. 
This study found that $\lamgas$ obeys a log-normal distribution similar to those found for $\lamh$ (and $\lamgas$) in simulations. 

\smallskip
At face value, the similarity between the distributions $P(\lamgal)$ and $P(\lamh)$ may naively suggest that $\lamgal$ of a given galaxy reflects $\lamh$ of the host halo.
However, one should note that the two spin parameters, $\lamgal$ and $\lamh$, are quantities that refer to different spatial and temporal scales -- the former mostly represents the sAM of the {\it inner} region and of recently arrived gas, while the latter represents the sAM of dark matter within the {\it whole} virial radius and is an integral over accretion throughout history. 
Generally, mass accreted more recently into the halo has higher sAM than the mass accreted earlier.
For example, at the virial radius of a halo, the spin parameter of the dark matter and cold gas streams are of the order of 0.1-0.2 in simulations \citep{danovich15}, significantly higher than the spin of the overall halo.
Therefore, the fact that the distributions of $\lamgal$ and $\lamh$ are similar, actually hints against AM conservation. 
 
\smallskip
\cite{danovich15} used cosmological simulations to study the evolution of gas sAM as gas streams from the cosmic web and feeds high-z galaxies. 
They described this evolution in four stages, as summarized in a cartoon in their Figure 20. 
They first revealed that outside the haloes, since the gas streams are thinner than the dark matter streams, they have higher quadrupole moments of the inertia tensor. 
Therefore, upon entering $\Rv$, the sAM of cold gas is highly correlated to, but $\sim1.5$ times higher than, the sAM of the incoming dark matter.
Inside $\Rv$, gas first maintains its AM in the outer halo while the incoming dark matter mixes with the existing halo of lower spin; then instreaming gas loses AM due to dissipation and torques from the inner galactic disc. 
Finally, torques and mass exchanges within the gaseous discs cause further changes in the disc sAM.
That is, overall, the gas gains and loses sAM with respect to the dark matter at different stages due to various mechanisms. 
Any mechanism that causes AM exchange or difference between the inner and outer parts of a halo, or any time variation of the AM supply in the cosmic web, can cause $\fj$ to deviate from unity. 
Since these processes differ in strength from galaxy to galaxy and from time to time, $\fj$ is expected to be stochastic, and the initial tight correlation between the sAM of incoming gas and dark matter is expected to be weakened or possibly smeared out.
If $\fj$ is stochastic or it is anti-correlated with $\lamh$, the simple recipe for galaxy size as given by \equ{Re1} is problematic. 

\smallskip
\cite{kravtsov13} compiled a sample of nearby galaxies of mixed morphologies and inferred the relation between galaxy size and halo virial radius using abundance matching.
He found that the data are scattered about a linear relation, $\Re\simeq A\Rv$, with $A=0.015$ on average, and a scatter $\sigma_{\log A}\approx0.25$dex.
Interestingly, not only the median of the proportionality factor is of the order of what is expected for $\lamh$, but also the scatter is similar to $\sigma_{\log\lamh}$.
\cite{huang17} and \cite{somerville17} extended the study of the kind of \cite{kravtsov13} to $z=0$-3.
They confirmed the form of the linear relation, but reported slightly larger proportionality factors, with noticeable dependences on redshift. 
\cite{somerville17} showed in detail that the dispersion in the conditional size distribution in stellar mass bins is in excellent agreement with the simple ansatz $\Re \propto \lamh \Rv$. 
However, as they point out, this does not prove that there is a strong one-to-one correlation between $\Re$ and $\lamh \Rv$.
Given that $\lamh$ is expected to be almost constant over time and as a function of mass, the dependence of the proportionality factor on mass and redshift hints that other parameters in addition to $\lamh$ must play a role in determining galaxy size.

\smallskip
These concerns cast doubt on the simple assumptions of a strong one-to-one correlation between the spins of a galaxy and its host halo, and the role of halo spin in determining the galaxy size. 
This motivates us to examine these issues directly in cosmological hydrodynamical simulations. 
We use two suites of simulations with very different subgrid physics for this study.
If no correlation between $\lamgal$ and $\lamh$ is found, we will look for a revised recipe for predicting galaxy size in semi-analytic models.

\smallskip
The outline of this paper is as follows.
In \se{method}, we describe the simulations.
In \se{dist}, we compare the distributions of spin of different components -- cold gas, stars, and dark matter halo.
In \se{correlation}, we characterize the correlations between galaxy spin and halo spin.
In \se{correlation2}, we present the correlations between gas spin and stellar spin. 
In \se{evolution}, we explore the possible effects of wet compaction (a compact star forming phase that many galaxies undergo at high-$z$) and mergers on the $\lamgal$-$\lamh$ correlation (or the lack therefore). 
In \se{size}, we study the relation between galaxy size and halo virial radius. 
In \se{discussion}, we compare our results to previous studies. 
In \se{conclusion}, we summarize our conclusions.

\section{Method}
\label{sec:method}

\smallskip
In this study, we use two suites of zoom-in hydro-cosmological simulations, VELA and NIHAO, of different hydro-gravitational solver, resolution and different recipes for cooling, star formation, and stellar feedback.

\subsection{VELA}

\smallskip
VELA is a suite of zoom-in hydro-cosmological simulations of 34 moderately massive central galaxies. 
It ultilizes the Adaptive Refinement Tree (ART) code \citep{krav97,krav03}, which follows the evolution of a gravitating $N$-body system and the Eulerian gas dynamics using an adaptive mesh. 
At the subgrid level, the code incorporates the key physical processes relevant for galaxy formation, including gas cooling by atomic hydrogen and helium as well as by metals and molecular hydrogen, photoionization heating by the UV background with partial self-shielding, star formation, stellar mass-loss, metal enrichment of the interstellar medium, and stellar feedback. 

\smallskip
Star formation is stochastic in cells with gas temperature $T < 10^4$ K and densities $n_{\rm H} > 1$ cm$^{-3}$ , at a rate consistent with the Kennicutt-Schmidt law \citep{Kennicutt98}.
Supernovae and stellar winds are implemented by local injection of thermal energy at a constant rate, as in \cite{ck09}, \cite{cdb10}, and \cite{ceverino12}. 
Radiative stellar feedback is implemented with no significant infrared trapping, motivated by \cite{dk13}, as described in \cite{ceverino14}.

\smallskip
The simulation adopts the Wilkinson Microwave Anisotropy Probe 5 cosmological parameters ($\omm= 0.27$, $\oml = 0.73$,  $\omb = 0.045$, h = 0.7, $\sigma_8 = 0.82$) \citep{WMAP5}. 
The dark matter particle mass is $8.3 \times 10^4 \Msun$, in the zoom-in region, and star particles have a minimum mass of $10^3 \Msun$. 
The AMR cells are refined to a minimum size in the range 17-35 pc at all times in the dense  regions. 
The force resolution is two grid cells, as required for computing the gradient of the gravitational potential. 
Each cell is split into eight cells once it contains a mass in stars and dark matter higher than $2.6 \times 10^5 \Msun$, equivalent to three dark matter particles, or once it contains a gas mass higher than $1.5 \times 10^6 \Msun$. 
The output timesteps that are analyzed are uniform in scale factor, with $\Delta a = 0.01$.

\smallskip
Among the 34 galaxies, 28 galaxies reach $z=2$, with the majority reaching $z=1$ and three of them reaching $z=0.8$. 
The VELA galaxies are selected to be systems that, at $z=1$, show no on-going major mergers, and are in the mass
\footnote{Throughout the paper, halo mass is defined as the total mass within a sphere if radius $\Rv$ that encompasses an overdensity of $\Delta_{\rm vir}(z)$ times the critical density of the Universe \citep{bn98}.} 
range of $\Mv=2\times10^{11}$--$2\times10^{12}\Msun$, about a median of $4.6\times10^{11}\Msun$.
At $z = 2$, the sample span the halo mass range of $(1$--$9)\times10^{11}\Msun$, and the stellar mass range of $(0.2$--$5.7)\times10^{10}\Msun$.
If they were to evolve to $z=0$, their mass range would bracket the Milky-Way mass.
More details concerning the simulations are presented in \cite{ceverino14} and \cite{zolotov15}.

\subsection{NIHAO}

\smallskip
We also use a subset of the Numerical Investigation of a Hundred Astrophysical Objects (NIHAO) project \citep{wang15}, consisting of 13  central galaxies that are Milky-Way sized or slightly more massive at z=0 [$\Mv = 7 \times 10^{11}$ - $3 \times 10^{12} \Msun$], evolved using the SPH code Gasoline 2.0\citep{wadsley17}.
The code includes a subgrid model for turbulent mixing of metals and energy \citep{wadsley08}, ultraviolet heating, ionization and metal cooling \citep{shen10}. 
Star formation and feedback follows the model used in the MaGICC simulations \citep{stinson13}, adopting a threshold for star formation of $n_{\rm H} > 10.3$ cm$^{-3}$.
Stars feed energy back into the interstellar medium via blast-wave supernova feedback \citep{stinson06} and early stellar feedback from massive stars. 

\smallskip
The simulations are run in a flat $\Lambda$CDM cosmology with parameters from the Planck Collaboration \cite{planck15}
 ($\omm = 0.3175$, $\oml = 0.6824$, $\omb  = 0.0490$, $h=0.671$, $\sigma_8 = 0.8344$, $n = 0.9624$).
Particle masses and force softenings are chosen to resolve the mass profile to below 1 per cent of the virial radius at all masses, ensuring that galaxy half-light radii are well resolved. 
For the 13 galaxies that we use, the particle mass is $1.735 \times 10^6\Msun$ for dark matter, and $3.166 \times 10^5\Msun$ for gas.
The force softening length is 931.4pc for dark matter, and 397.9pc for baryons (comoving).
The output is uniform in cosmic time, with $\Delta t\simeq 215$Myr, approximately $\Delta a \simeq 0.014$ at $z\sim1$.

\begin{figure}
\includegraphics[width=0.48\textwidth]{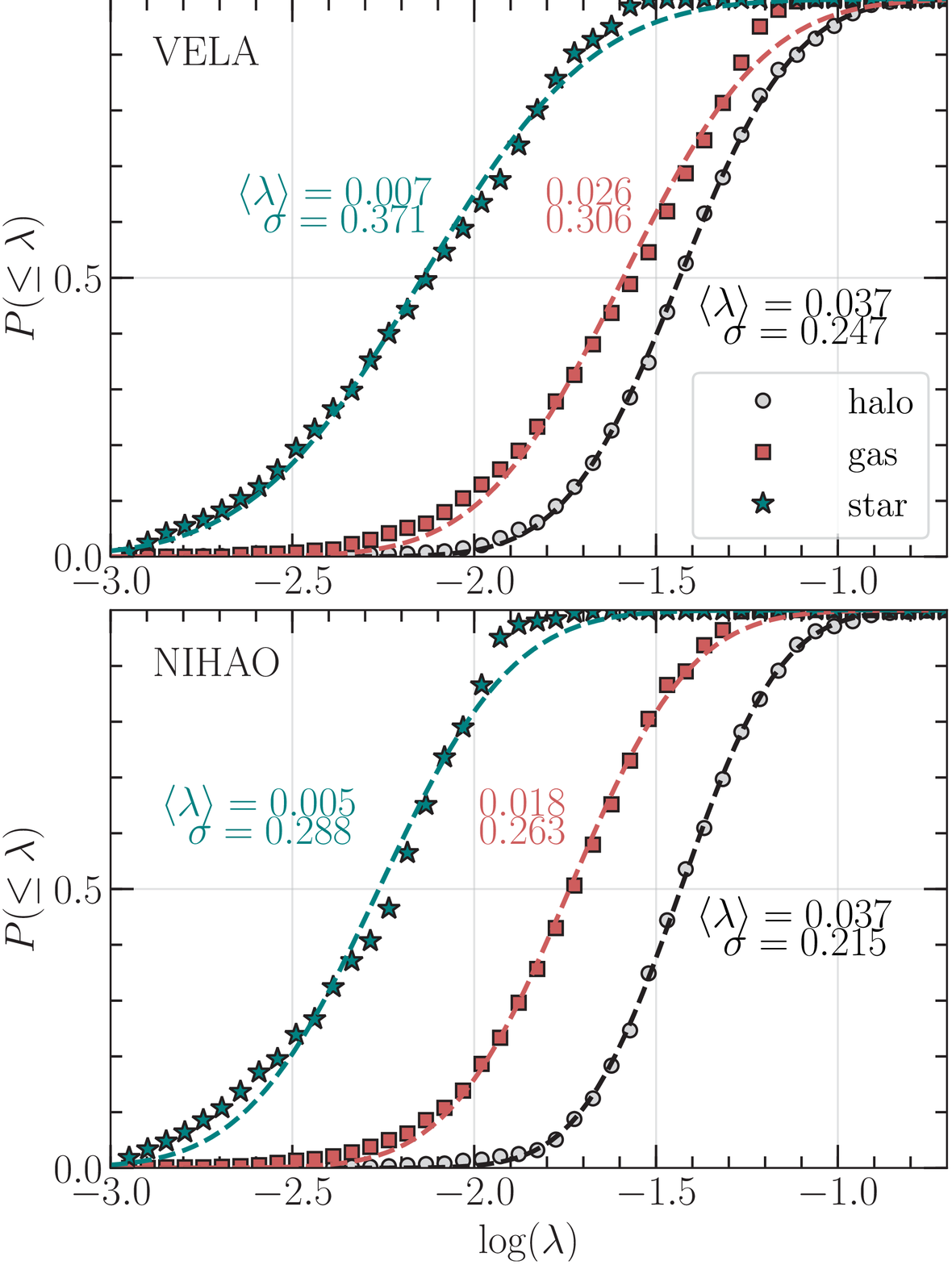} 
\caption{
Cumulative distributions of the spin for host halo (within $\Rv$), cold gas (within 0.1$\Rv$), and stars (within 0.1$\Rv$) of the VELA ({\it upper}) and NIHAO ({\it lower}) simulations. 
The lines represent the best-fit log-normal distributions. 
$\lamgas$ has a mean value slightly lower than $\lamh$, and exhibits a marginally larger scatter, while $\lams$ is much lower than the other components and follows a much broader distribution.
The NIHAO $\lams$ and $\lamh$ distributions are similar to those of VELA; while $\lamgas$ in NIHAO is 0.15 dex lower.
}
\label{fig:spinCDF} 
\end{figure}
\begin{figure*}
\includegraphics[width=\textwidth]{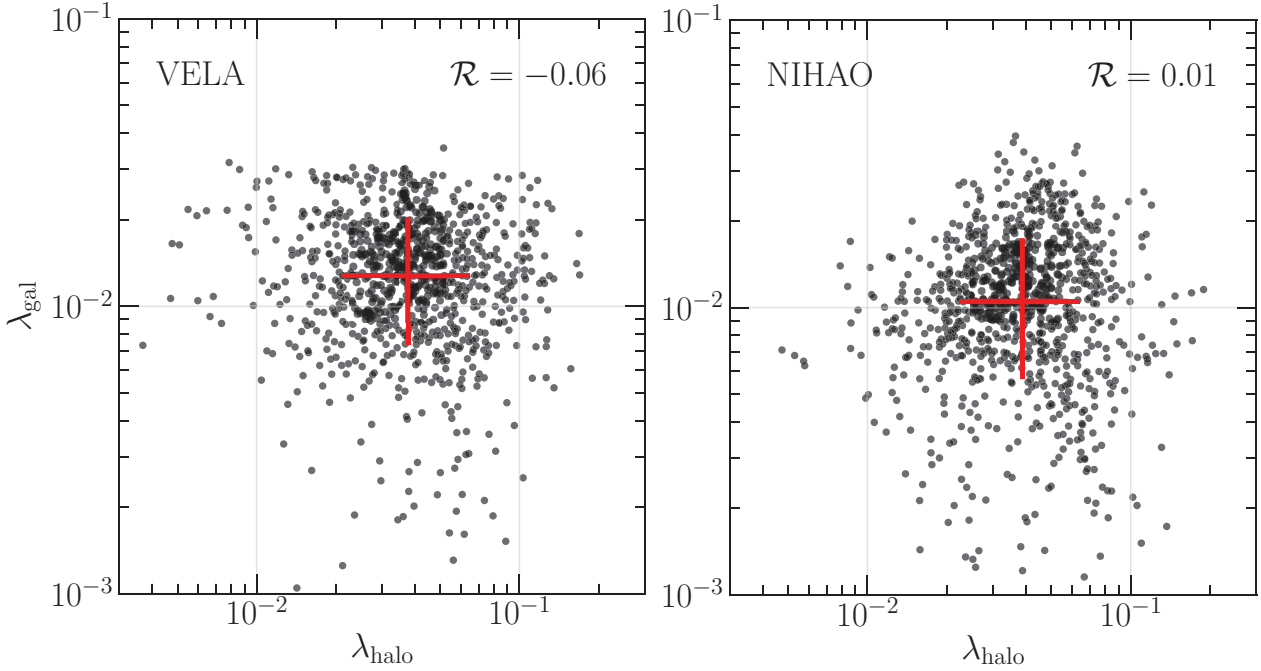} 
\caption{
The spin of a galaxy (stars+cold gas within 0.1$\Rv$) versus the spin of its host halo (DM within $\Rv$), for the VELA ({\it left}) and NIHAO ({\it right}) simulations.
Each dot is {\it a galaxy at one snapshot}. 
Here we have corrected for systematic dependence on halo mass or redshift (see text).
The cross marks the median and the 16th and 84th percentiles. 
The Pearson correlation coefficient $\mathcal{R}$ is quoted at the upper right corner.
In both simulations, there is negligible correlation between $\lamgal$ and $\lamh$.
}
\label{fig:correlation} 
\end{figure*}
%

\subsection{Comments}

\smallskip
The NIHAO galaxies reach $z=0$, while more than half of the VELA galaxies reach $z=1$, some reaching $z=0.8$.
Both suites consist of central galaxies, selected to be far from larger neighbours throughout their assembly history, and would be $\sim L_\star$ or sub-$L_\star$ galaxies at $z=0$, if left evolving in isolation.

\smallskip
In what follows, in order to probe potential redshift dependence, we bin the snapshots into three redshift ranges, $z=$0-0.8, 0.8-2, and 2-7. 
To achieve better statistics, we consider {\it each galaxy at each snapshot} as an {\it independent} measurement. 
The median halo masses of the NIHAO snapshots are $\log(\Mv/\Msun)=11.82^{+0.21}_{-0.17}$, $11.59^{+0.29}_{-0.27}$, and $10.86^{+0.49}_{-0.43}$, in the three redshift ranges, respectively. 
Here the upper and lower limits indicate the 16th and 84th percentiles. 
The median halo masses of the VELA snapshots are $\log(\Mv/\Msun)=11.51^{+0.32}_{-0.25}$ and  $10.99^{+0.45}_{-0.66}$ respectively, for $z=$0.8-2, and 2-7, respectively. 

\smallskip
The two suites differ in many subgrid recipes, and in particular in numerical resolution and in the implementation of stellar feedback. 
The VELA simulations have relatively weak stellar feedback.
The stellar-to-halo mass ratio in the VELA simulations is in ballpark agreement (considering the scatter in these relations) with the results of abundance matching relations from \cite{behroozi13a} and \cite{moster13}.
The stellar feedback of the NIHAO simulation is much stronger than VELA, and is tuned to match the $z=0$ stellar mass versus halo mass relation from abundance matching.
The two simulations are therefore complementary in terms of revealing potential dependence of the results on the feedback strength. 
Neither VELA or NIHAO has included AGNs. 
The effects of AGNs may start to become important for $L_\star$-galaxies.
The most massive end of our sample may be affected and therefore should be taken with caution.

\smallskip
The NIHAO galaxies are mostly on the star formation main sequence, while the VELA simulations exhibit more diverse sSFR.
Very few galaxies in our sample are completely quenched.
There have been a plethora of studies discussing the correlation between morphology and angular momentum loss, and this is {\it not} the focus of this paper -- we generally do not distinguish early-type and late-type galaxies in the following, unless there is a significant morphology-dependence in the results. 
We verify though that our main results hold if we only use the galaxies on or off the main sequence.

\subsection{Measuring Spin}

\smallskip
Galaxy centers are defined as follows.
For the VELA simulations, we identify the cell of the highest baryonic density, draw a sphere of 1kpc around it, and take the center-of-mass (CoM) of the stars in the 1kpc-sphere as the center position, $\mathbf{r_0}$. 
The CoM velocity of these stars is taken as the bulk velocity, $\mathbf{v_0}$, which is then used to define the rest frame for calculating the AM.

\smallskip
For the NIHAO galaxies, we take the center of the host halo as given by the {\tt AHF} halo catalog \citep{knollmann09} as an initial guess, and run a shrinking-sphere algorithm on the stars within 5kpc from the initial center until the shrinking sphere contains 50 particles.
We take the outcome of the shrinking-sphere algorithm as $\mathbf{r_0}$, and the CoM velocity of the stars in the 5kpc-sphere as $\mathbf{v_0}$.
The size of the sampling sphere, 1kpc and 5kpc for VELA and NIHAO, respectively, is chosen such that it is well above the spatial resolution of the simulation and is significantly smaller than $\Rv$ (to avoid contaminations from massive satellite galaxies). 

\smallskip
With visual inspections of the projected density maps of dark matter, gas, and stars, we find that the centers defined above are sensible -- although they are basically the locations of the highest stellar mass density, they overlap with the positions of the highest dark matter and gas mass densities within a couple of softening lengths, even during major mergers.
We also verify that our main results are not sensitive to the definition of centers: using sampling spheres of 0.01$\Rv$ or using the centers of dark matter instead of stars yields statistically indistinguishable results.

\smallskip
The spin parameters of cold\footnote{Density $n_{\rm H} > 1$ cm$^{-3}$ and temperature $T<10^4$K.} gas ($\lamgas$) and stars ($\lams$) are defined by \equ{spin}, with the sAM $j$ measured within 10\% of the virial radius $\Rv$.
Dark matter halo spin is defined within $\Rv$. 
The AM $J$ is defined about the origin $(\mathbf{r}_0,\mathbf{v}_0)$.
Throughout, we consider a {\it galaxy} to be the stars and cold gas within 10\% of the virial radius of the host halo, and usually present the galaxy spin $\lamgal$ measured from stars and cold gas combined, unless the spins of the stars and gas show different trends. 

\smallskip
There is a corresponding dark-matter-only (DMO) simulation for each NIHAO galaxy.
The DMO simulations adopt the same initial conditions as the hydro ones, and replace the gas particles with dark matter particles of the same mass.
We measure the dark matter properties of the simulations with baryons, and mention the check on the DMO outputs when necessary. 

\section{Distribution of spin for different components}
\label{sec:dist}

\smallskip
\fig{spinCDF} shows the cumulative distributions of the spins of the cold gas, stars and dark matter halo, for all snapshots after $z=7$.
The spin parameters of the different components are all well-described by log-normal distributions.
Halo spin obeys a log-normal distribution with a mean of $\langle\lamh\rangle=0.037$, and a standard deviation of $\sigma_{\log\lambda}\simeq0.2$-0.25, very similar to those found in $N$-body simulations (e.g. \citealt{bullock01j}, \citealt{bett07}; \citealt{mc11}; \citealt{somerville17}; \citealt{lee17a}).

\smallskip
In both simulations, the stellar spin is $\sim5$ times lower than the cold gas spin, which, in turn, is slightly lower than the halo spin. 
The cold gas spin is only 30-50\% lower than halo spin, with $\langle\lamgas\rangle=0.026$ (VELA) and 0.018 (NIHAO). 
This indicates that the radial specific angular momentum profile of the gas is higher than that of the dark matter, to the extent of $j_{\rm gas}(0.1\Rv)$ being comparable to $j_{\rm dm}(\Rv)$.
Indeed, as discussed in \citet{danovich15}, the gas streams have higher sAM than the dark matter streams at accretion to $\Rv$ within a factor of two, but as they reach the inner halo, they are spinned down by the torque from the galactic disc. 
The scatter of $\lamgas$ is somewhat larger than that of $\lamh$, by $\sim$ 0.05dex in both cases.

\smallskip
The spin of cold gas ($\lamgas$) and of stars ($\lams$) of the NIHAO simulations are systematically lower than those of VELA . 
There can be several possible reasons. 
First, as to the difference in gas spin, the gas properties are different between the two suites.
Due to the relatively high density threshold of star formation, the gas in the NIHAO simulations can condense to higher densities than that in VELA before turning into stars.
Due to the stronger stellar feedback, NIHAO galaxies have stronger exchange of AM between the cold galaxy range and the hot halo.
Second, wet compaction, a process that high-$z$, stream-fed galaxies generally undergo, is efficient at raising $\lamgas$ (see \se{compaction} for more details). 
The strength of compaction is expected to be sensitive to numerical resolution and feedback, so the NIHAO galaxies, of much coarser resolution and stronger feedback than VELA, generally show much weaker compactions.
Finally, artificial losses of AM can occur in SPH simulations with limited resolution \citep{kaufmann06}. 
With a force softening length of a few hundred parsecs, and gas particle mass of a few times $10^5\Msun$, the NIHAO simulations could carry non-negligible numerical AM loss.  

\section{Spins of galaxy versus halo}
\label{sec:correlation}

\begin{figure*}
\includegraphics[width=1.05\textwidth]{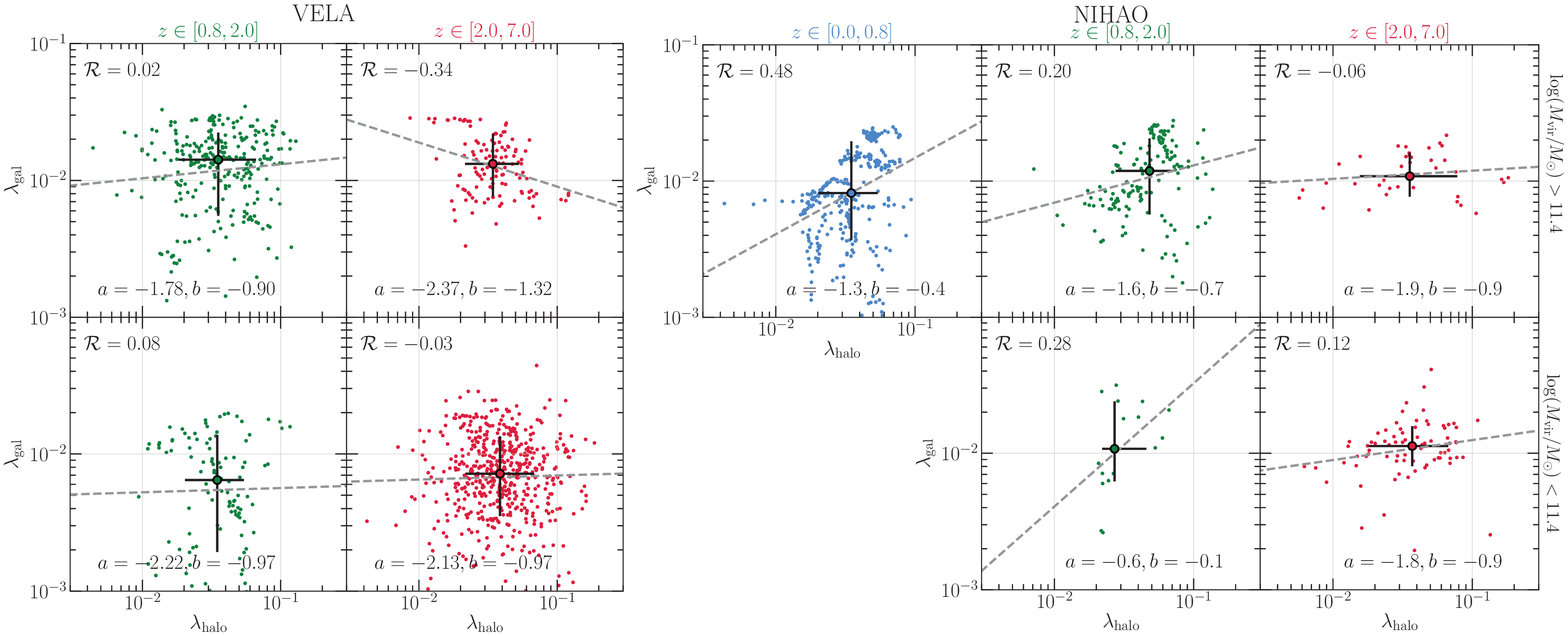} 
\caption{
The spin of a galaxy (stars+gas within 0.1$\Rv$) versus the spin of its host halo (DM within $\Rv$), in different bins of halo mass ({\it upper}: $\Mv>10^{11.4}\Msun$, {\it lower}:$\Mv<10^{11.4}\Msun$) and redshift, in the VELA simulation ({\it left}) and in the NIHAO simulation ({\it right}).
Each dot is {\it a galaxy at one snapshot}. 
The big circles mark the medians with error bars indicating the 16th and 84th percentiles.
The Pearson correlation coefficient $\mathcal{R}$ is quoted at the upper left corner.
The solid lines are linear regression of the form of $\log\lamgal = a + (1+b)\log\lamh$, with the best-fit parameters indicated.
The VELA simulation exhibits negligible correlation throughout all the $\Mv$ and $z$ bins.
$\lamgal$ is higher by a factor of $\sim$2 in systems with $\Mv > 10^{11.4}\Msun$  or equivalently post compaction (see text in \se{compaction}). 
In the NIHAO simulation, a weak correlation emerges at $z\la1$.
}
\label{fig:correlation_Mzbins} 
\end{figure*}

\smallskip
In this section, we characterize the correlation of the {\it amplitudes} of the spins of the galaxies and their host haloes, as well as the alignment of the spin {\it vectors}.

\subsection{Spin amplitude}

\smallskip
\fig{correlation} presents galaxy spin versus host halo spin for all the galaxies at all snapshots at $z<7$.
Here, we have tried to remove potential trends with halo mass or redshift as follows. 
First, the sample is binned into three redshift ranges, $z=0$-0.8, 0.8-2, and 2-7, and two halo mass ranges, $\Mv>10^{11.4}\Msun$, and $<10^{11.4}\Msun$. 
We then calculate the median $\lamgal$ and $\lamh$ for each mass and redshift bin, and for the whole sample.
Finally, in the $\lamgal$-$\lamh$ plane, the data points of different bins are shifted by the offset between the median point of the corresponding bin and that of the whole sample. 
As can be seen, there is almost no correlation between $\lamgal$ and $\lamh$ in either NIHAO or VELA.

In fact, as shown in \fig{correlation_Mzbins}, there is negligible correlation in almost every $(\Mv,z)$ bin at $z\ga0.8$ --
the Pearson correlation coefficient $\mathcal{R}$ seldom exceeds $0.3$.
There seems to be a correlation emerging in the lowest redshift bin in the NIHAO simulations, but it is still weak with $\mathcal{R}=0.48$.

\smallskip
\cite{danovich15} showed that the sAM of gas and dark matter are strongly correlated at virial crossing,
\footnote{In fact, \cite{danovich15} found $\lamgas \simeq $1.5$\lamdm$, with both $\lamgas$ and $\lamdm$ measured at $\ga\Rv$. 
Gas spin at accretion is slightly higher, due to the higher quadrupole moment resulting from the early dissipative contraction of gas into the central cords of the thick dark matter filaments.}
therefore, the lack of correlation between $\lamgal$ and $\lamh$ means that the spin of baryons evolves differently with respect to that of the dark matter, inside the virial radius. 
That is, the angular momentum retention ratio, $\fj\equiv\lamgal/\lamh$, must deviate from a constant of order unity, and vary from one galaxy to another and from time to time. 
We expect $\fj$ to depend systematically on $\lamh$ -- any mechanism that causes an anti-correlation between $\fj$ and $\lamh$ can most efficiently erase the initial correlation between the sAM of baryons and dark matter in the cosmic web.
Assuming for simplicity that the anti-correlation is parametrized by $\fj \propto \lamh^b$ with a negative $b$, one can write
\be
\label{eq:regression}
\log\lamgal = a + (1+b)\log\lamh,
\ee
where $a$ is the zero point of the relation, and $(1+b)$ is also a measure of the correlation strength between $\lamgal$ and $\lamh$ (in addition to the Pearson $\mathcal{R}$), that ranges from proportionality ($1+b=1$) to no correlation ($1+b=0$).

\smallskip
The dashed lines in \fig{correlation_Mzbins} are linear regressions of the form of \equ{regression}.
Clearly, there is always $-1\la b<0$ across all the redshift and halo mass bins.
Hence, to smear out an initial correlation between the sAM of baryons and dark matter in the cosmic web, some mechanisms operate inside the halo such that initially high-$\lamh$ systems end up having lower-$\lamgal$; and perhaps also that initially low-$\lamh$ systems end up with higher $\lamgal$.
We discuss two possible mechanisms of this sort, wet compaction and mergers, in \se{evolution}.

\smallskip
The left-hand panels of \fig{correlation_Mzbins} show that, in the VELA simulations, $\lamgal$ is higher in more massive haloes.
As will be discussed in more detail in \se{compaction}, this is basically a manifestation of galaxy compaction -- galaxies above the mass threshold $\Mv\sim10^{11.4}\Msun$ are typically post compaction, where the sAM is higher due to an extended ring that has formed from newly accreted gas. 

\smallskip
By inspecting the spin of stars and cold gas separately, we verify that the results are qualitatively the same, with a weak to null correlation between either baryonic component within $0.1\Rv$ and the dark matter halo within $\Rv$.

\subsection{Alignment}

\begin{figure} 
\includegraphics[width=0.5\textwidth]{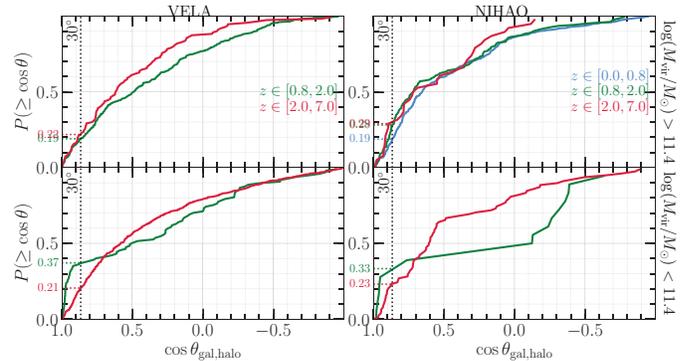} 
\caption{
Cumulative distribution of the cosine of the angle between the angular momentum vectors of the galaxy (stars + cold gas within 0.1$\Rv$) and that of the host halo (dark matter within $\Rv$), for VELA ({\it left}) and NIHAO ({\it right}) galaxies in different redshift and halo mass bins.
Dotted lines mark the fraction of systems with $\theta<30^\circ$.
In VELA, the median $\cos\theta=0.53$-0.67, and seems to decrease (i.e., the alignment becomes marginally worse) at lower redshift.
The NIHAO galaxies exhibit better alignment than VELA, with a similar, weak redshift trend. The lower-$\Mv$ bin of NIHAO likely suffers from small-number statistics.
}
\label{fig:alignment} 
\end{figure}

\smallskip
Although the amplitudes of spin are barely correlated, are the spin vectors of baryons and dark matter randomly oriented? 
To answer this question, we plot in \fig{alignment} the cumulative distributions of $\cos\theta_{\rm gal,halo} = \jjgal \cdot \jjh / |\jjgal||\jjh|$ for different halo mass and redshift bins.

\smallskip
Generally, the median $\cos\theta_{\rm gal,halo}$ is in the range of 0.6-0.7. 
Approximately $40$\% ($20-30$\%) systems have $\cos\theta_{\rm gal,halo}>$ 0.71 (0.87), corresponding to an angle of $\theta_{\rm gal,halo}<45^\circ$ ($30^\circ$). 
At a given halo mass, the alignment becomes marginally weaker at later times.

\smallskip
Comparing the two simulation suites, for the more massive haloes, NIHAO exhibits a slightly better alignment, with the median $\cos \theta_{\rm gal, halo}=$0.65-0.72, while VELA shows a median of $\cos \theta_{\rm gal, halo}=$0.50-0.62 depending on redshift.
For the less massive cases, there seems to be a significant fraction with $\cos\theta\la0$.
We note though, that the NIHAO results at $\Mv<10^{11.4}\Msun$ suffer from small-number statistics, and therefore opt not to overinterpret them.

\smallskip
Not shown here, we find the alignment between cold gas and halo to be better than that between stars and halo. 
In particular, the NIHAO galaxies with $\Mv>10^{11.4}\Msun$ have a median $\cos\theta_{\rm gas, halo}$ of 0.71-0.73 and a median $\cos\theta_{\rm stars, halo}$ of 0.64-0.69 (depending on redshift weakly), compared to the corresponding VELA results of 0.59-0.68 and 0.48-0.62, respectively.
For VELA galaxies with $\Mv<10^{11.4}\Msun$, the medians of $\cos\theta_{\rm gas, halo}$ and $\theta_{\rm stars, halo}$ are 0.57-0.63 and 0.53-0.57, respectively.

\smallskip
It is intriguing that the amplitudes of spins are uncorrelated although the alignment of the spin vectors is relatively good. 
It may well be that because the gas streams and the stellar disc are generally coplanar \citep{danovich12}, the torques that cause angular momentum gain or loss do not randomize the directions of the spin vectors.

\subsection{Spin of galaxy versus inner halo}

\begin{figure} 
\includegraphics[width=0.45\textwidth]{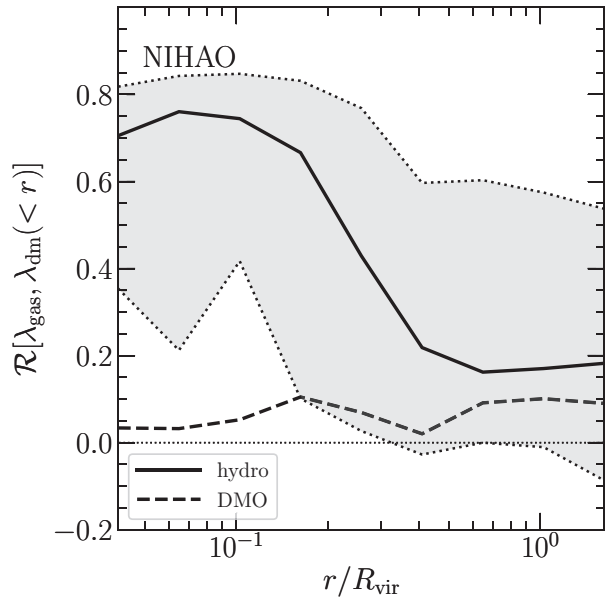} 
\caption{
Correlation between the spin of cold gas measured within $0.1\Rv$ and the spin of dark matter within radius $r$, as a function of radius $r$, for the NIHAO galaxies.
The {\it solid} line represents the median result over the redshift range $z=0-7$. 
The dark matter spin is measured in the (fiducial) hydrodynamical simulation, with the shaded region bracketed by the dotted lines representing the 16th and 84th percentiles. 
The {\it dashed} line represents the median result, for which the dark matter spin is measured in the dark-matter-only (DMO) simulation that is complementary to the NIHAO simulation and that uses the same initial condition and replaces the gas particles in the fiducial NIHAO simulation with dark matter particles of equal mass.
Gas spin is strongly correlated with the spin of dark matter out to $\sim0.2\Rv$ if the dark matter spin is measured in the hydrodynamical simulation; while there is negligible correlation between gas spin and the dark matter spin measured in the DMO simulation.
}
\label{fig:correlation_inner} 
\end{figure}
\begin{figure*}
\includegraphics[width=\textwidth]{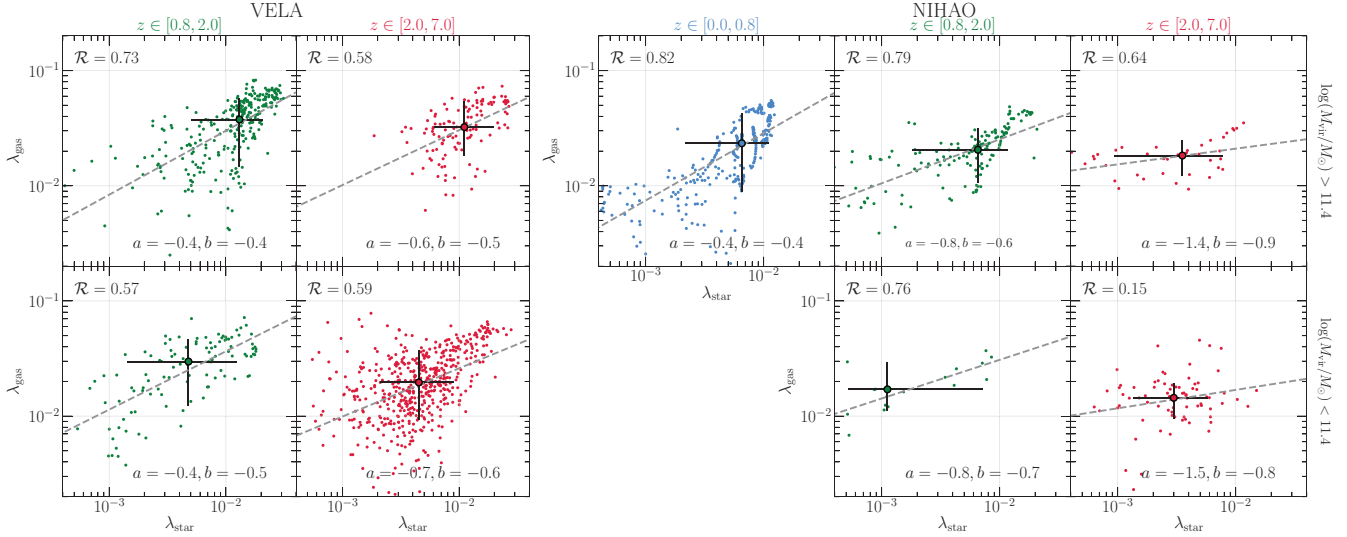} 
\caption{
The spin of cold gas versus the spin of stars, for VELA ({\it left}) and NIHAO ({\it right}) galaxies in different bins of halo mass and redshift.
In both VELA and NIHAO, the spins of the baryonic components are strongly correlated, and the correlation is stronger at later times. 
On average, the spin of either baryonic component is higher in more massive (post compaction) systems.
}
\label{fig:correlation_gasstar} 
\end{figure*}

\smallskip
The correlation between the galaxy spin and the {\it whole} halo spin is weak, but the two parameters sample very different spatial scales. 
It may well be that the spin of the dark matter in the inner part of the halo, where the galaxy dwells, correlates with the galaxy spin better.
To check this, we focus on the NIHAO galaxies, and measure the Pearson correlation coefficient between the gas spin, $\lamgas$, and the spin of the dark matter within radius $r$, $\lamdm(<r)$, at each snapshot, looking for a radius within which the dark matter spin is a good proxy of the gas spin.

\smallskip
The result is shown in \fig{correlation_inner}.
On average, there is a strong correlation ($\mathcal{R}\sim0.7$) out to $r\sim0.2\Rv$. 
Beyond $0.2\Rv$, the correlation drops quickly to negligible. 
The scatter of the correlation profile largely reflects redshift dependence: the correlation is  stronger at lower redshift, as already hinted in the right-hand panels of \fig{correlation_Mzbins}.

\smallskip
From the point of view of semi-analytic modelling, we are more interested to know whether or not one can use halo properties measured in $N$-body simulations to predict galaxy properties. 
Therefore, to test if $\lamdm(<0.2\Rv)$ is an adequate galaxy spin indicator, we repeat the exercise, re-measuring $\lamdm(r)$ in the dark-matter-only (DMO) simulation complementary to the NIHAO sample. 
The DMO simulations adopt the same initial conditions as the fiducial NIHAO sample, and replace the gas particles with dark matter particles of the same mass. 
As shown by the dashed line in \fig{correlation_inner}, the $\lamdm(<r)$ measured in the DMO simulation barely correlates with the gas spin, irrespective of radius $r$.
Hence, the DMO inner halo spin, $\lamdm(<0.2\Rv)$, is not eligible to be a proxy of galaxy spin as would be used in semi-analytic modeling, and baryonic processes influence the inner halo spin and cause an correlation.
What specific mechanisms cause this correlation is beyond the scope of this study. 
Speculatively, the same processes could simultaneously be related to the null correlation between the galaxy spin and the whole halo spin.

\section{Spins of gas versus stars}
\label{sec:correlation2}

\smallskip
Here we measure the correlation between the spins of stars and cold gas in the galaxy.
On one hand, a correlation is expected to reflect the fact that stars have formed from cold gas and that the accreted stars and gas might have suffered similar torques. 
On the other hand, the stars may reflect the spin of the gas at earlier times, which may be different from that of the newly accreted gas. 

\subsection{Spin amplitude}

\smallskip
As shown in \fig{correlation_gasstar}, for the two simulation suites, the spins of the baryonic components are correlated, with $\mathcal{R}\sim0.6$-0.8.
In terms of redshift trend, both simulations seem to show a mildly stronger correlation at lower redshift.
As for halo mass dependence, in the VELA sample, both $\lamgas$ and $\lams$ are higher by $\sim$50\% for systems with $\Mv>10^{11.4}\Msun$ than for $\Mv<10^{11.4}\Msun$.
NIHAO shows the same qualitative trend, although the baryonic spins in NIHAO are overall slightly lower than in VELA.
NIHAO again exhibits marginally stronger correlations than VELA in general. 

\subsection{Alignment}

\begin{figure}
\includegraphics[width=0.5\textwidth]{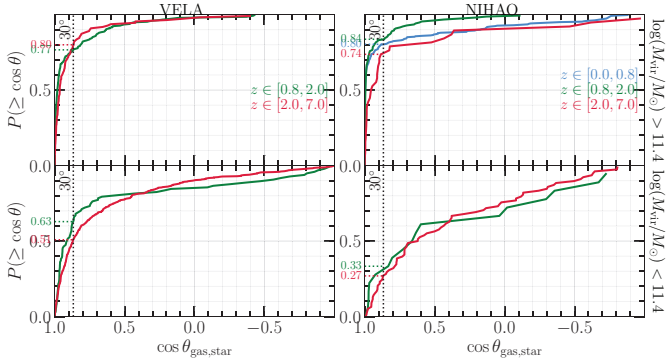} 
\caption{
Cumulative distribution of the cosine of the angle between the angular momentum vectors of cold gas and stars, both measured within $0.1\Rv$, for VELA ({\it left}) and NIHAO {\it right} galaxies in different redshift and halo mass bins.
The redshift and halo mass bins are the same as used in previous shows, as indicated.
Dotted lines indicate the fraction of systems with $\theta<30^\circ$.
In both simulation suites, the spin vectors of the baryonic components are well aligned. 
The alignment is better in more massive systems, and strengthens marginally at later times.
}
\label{fig:alignment_gasstar} 
\end{figure}

\smallskip
\fig{alignment_gasstar} shows the cumulative distributions of $\cos\theta_{\rm gas,star} = \jjgas \cdot \jjstar / |\jjgas||\jjstar|$.
The gas and stellar spin vectors are generally well aligned. 
The median $\cos\theta_{\rm gas,star}$ is 0.96-0.98 (0.96-0.99) in the VELA (NIHAO) simulation at $\Mv>10^{11.4}\Msun$, and is 0.88-0.92 in the VELA simulation at $\Mv<10^{11.4}\Msun$.
The gas and stellar spin vectors are better aligned in more massive systems, and there seems to be a weak trend that the alignment becomes marginally better at later times. 

\smallskip
We notice that a non-negligible fraction of galaxies have {\it counter-rotating} gas and stellar components, especially the less massive ones. 
In particular, the fraction of having $\theta_{\rm gas, stars} > 120^\circ$ is 5\% (9\%) and 0\% (3\%) in the VELA (NIHAO) simulation, for $\Mv<10^{11.4}\Msun$ and $>10^{11.4}\Msun$, respectively.

%


\section{Origin of the null correlation between galaxy and halo spins}
\label{sec:evolution}

\smallskip
In this section, we discuss two mechanisms that can cause $\fj$ to anti-correlate with $\lamh$, and speculate on other possible processes that can decouple $\lamgal$ from $\lamh$.

\subsection{Effect of compaction}
\label{sec:compaction}

\begin{figure*}
\includegraphics[width=\textwidth]{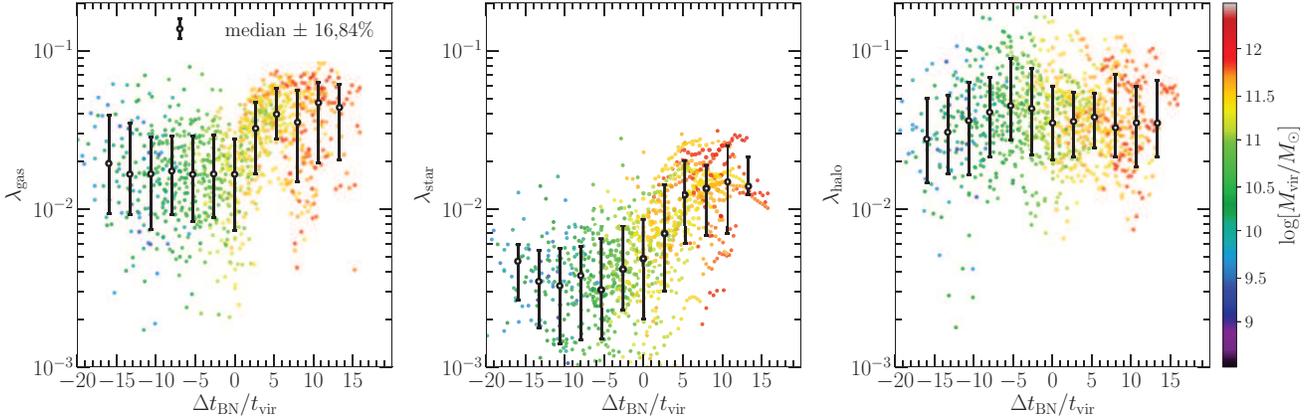} 
\caption{ The spin of cold gas, stars (within $0.1\Rv$), and dark matter halo (within $\Rv$) versus $\Delta\tbn/\tv$ in the VELA simulation, where $\Delta\tbn/\tv$ is the time to the BN snapshot in units of virial time.
The spins of baryons are rising after the compaction event, due to the formation of an extended ring, while $\lamh$ remains roughly constant.
Color marks halo mass, indicating that compactions in VELA generally occurs at $\ga10^{11-11.5}\Msun$. 
}
\label{fig:compaction} 
\end{figure*}

\smallskip
One possible origin for the anti-correlation between $\fj$ and $\lamh$ is the dramatic compaction event that most galaxies undergo.  
\cite{db14} argued analytically, and \cite{zolotov15} and \cite{tacchella16_prof} showed using simulations, that most galaxies undergo phases of dissipative gas contraction triggered by mergers or counter-rotating accretion streams, into compact, star-forming systems, termed ``blue nuggets" (BN).
These objects have been observed \citep{barro13,barro14_bn_rn,barro14_kin,barro15_kin,barro17_uni,barro16_kin}. 
Observationally, these compact star forming nuclei may be obscured by dust and are not necessarily blue in color.
In the VELA simulations, compaction triggers inside-out quenching once above a threshold mass -- stellar mass $10^{9.5-10}\Msun$ and halo mass $10^{11-11.5}\Msun$. 

\smallskip
Wet compaction tends to occur when the incoming material has low sAM \citep{db14}. 
During the subsequent blue-nugget phase, the central gas is depleted to star formation and the associated outflows. 
These outflows preferentially eject low-AM gas, while the new incoming gas with higher spin settles in an extended ring (e.g. Fig. 7 in \citealt{zolotov15}). 
As a result, the gas spin in the galaxy is sharply rising during the BN phase. 
This implies that with a low $\lamgas$ at halo entry (presumably reflecting low $\lamh$, \citealt{danovich15}), the galaxy, that undergoes compaction, ends up having a higher $\lamgas$, namely $\lamgal/\lamh>1$. 

\smallskip
The spin of stars exhibits a similar behavior.
During the BN phase, new stars form in the center following the compact gas, but after the BN phase the stellar effective radius is gradually growing, partly due to new stars that form in the outer ring with high AM and partly due to new ex-situ stars from minor mergers, likely with higher AM.

\smallskip
We illustrate this effect using the VELA simulations.
We detect the main BN phase by identifying the snapshot of the most prominent increase in the surface gas density within 1kpc (see Dekel et al. 2018, in prep., for more details), and investigate the evolution of the spins of different components before and after the main BN phase.
\fig{compaction} shows $\lamgas$, $\lams$ and $\lamh$ versus $\Delta t_{\rm BN}/\tv$, the time from the BN phase in units of virial time.
Clearly, compaction affects the spins of baryons as aforementioned, but not the spin of the dark halo.
Since compactions are more common at high-$z$, this, combined with the correlation of compaction and low $\lamh$ reported in \citeauthor{zolotov15}, explains the anti-correlation between $\fj$ and $\lamh$ at high $z$.
Compaction also generally occurs at $\Mv\simeq10^{11-11.5}\Msun$, explaining the higher $\lamgal$ in more massive haloes as we have seen in the left-hand panels of \fig{correlation}.
We caution though that the post-compaction ring formation is hypothetical. 
The suppression of star formation by AGN feedback may become important for post-compaction systems and turn BNs into compact quiescent systems (``red nuggests'') quickly. 
Without AGNs, the VELA simulations may over-estimate the significance of the post-compaction ring phase. 

\subsection{Effect of mergers}
\label{sec:merger}

\smallskip
Another possible source for the anti-correlation of $\fj$ and $\lamh$ is the variations of spin during a major merger.
\footnote{Since the compaction is in many cases associated with a merger, the two may affect $\fj$ simultaneously.}
The accretion of a large satellite makes $\lamh$ rise, as the spin is temporarily dominated by the orbital angular momentum of the merging dark matter haloes. 
The galaxy spin $\lamgal$ is temporarily unaffected unless the satellite survives the tidal disruption and reaches the central baryonic range.
In such case, within a couple of halo dynamical times, $\lamh$ relaxes to its normal value as some of the high AM material ends up beyond $\Rv$ \citep{lee17b}, but $\lamgal$ rises, as the orbital angular momentum of the baryonic components now dominates the galaxy spin.
This two-phase process is expected to introduce an anti-correlation between $\fj$ and $\lamh$. 

\begin{figure}
\includegraphics[width=0.45\textwidth]{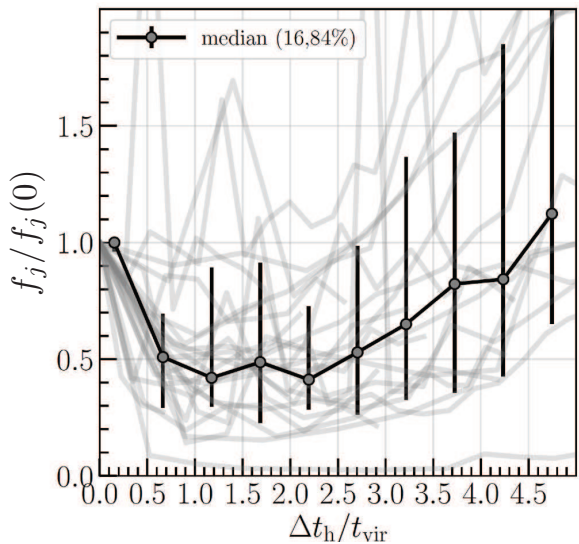} 
\caption{ The ratio $\fj\equiv \lamgal/\lamh$ with respect to $\fj$ at the moment of a massive halo merger (see text) as a function of the time after the merger, $\Delta\thm/\tv$, for the NIHAO galaxies.
Thin grey lines are individual cases; the thick line with error bars indicate the median and 16 and 84 percentiles. 
$\fj$ decrease immediately after halo merger and start to recover after $\sim$2 virial times, both phases giving rise to an anti-correlation between $\fj$ and $\lamh$.
}
\label{fig:merger} 
\end{figure}

\smallskip
We illustrate this effect using the NIHAO simulations. 
We detect halo mergers as mass increments of the main progenitor more than 10\% (i.e., major and minor mergers). 
\footnote{A mass ratio of 1:10 between the satellite and the central is chosen such that 1) orbit decay due to dynamical friction is efficient; 2) the time interval between two mergers is long enough to allow for the characteristic evolution pattern of spins as described above.}
If there are multiple detections of mergers within two halo dynamical times, $\tv\equiv\Rv/\Vv$, we only keep the earliest one, to avoid double counting the re-accretion of splashback satellites.
This gives us a clean, but not necessarily complete sample of massive mergers. 
\fig{merger} shows the spin ratio $\fj$, with respect to that at the moment of a halo merger $\fj(0)$, as a function of the time after halo merger, $\Delta\thm/\tv$. 
Clearly, $\fj$ drops abruptly upon the accretion of a large satellite, and starts to recover after $\sim2\tv$.
We verify that VELA simulations also show the decrease-and-recover behavior of $\fj$.
As with compaction, massive mergers are also more frequent at high-$z$, helping to partly explain the anti-correlation between $\fj$ and $\lamh$ at high $z$.
We caution that both effects are more common at higher masses, and the low mass bin perhaps requires other explanations.

%

\subsection{Other reasons for the lack of correlation}

\smallskip
While the two mechanisms discussed above generate some anti-correlation, we learn that they are not enough for explaining the full effect.
For example, removing the post-halo-merger snapshots within 4$\tv$ in the VELA simulation results in a positive, but rather weak correlation between $\lamgal$ and $\lamh$, with $\mathcal{R}\approx0.3$.
The two mechanisms are tightly related -- in fact, about 40\% of compactions are preluded by massive mergers, and the rest are associated with minor mergers, disk instabilities, counter-rotating streams, or other mechanisms (Dekel et al. 2018, in preparation).
Here we speculate on a few other possible processes that may smear out the $\lamgal$-$\lamh$ correlation but do not necessarily cause an anti-correlation between $\fj$ and $\lamh$.
We refer interested readers to \cite{danovich15}, for a comprehensive discussion of different stages of AM build up for galaxies at high-$z$. 

\smallskip
First, $\lamgal$ and $\lamh$ are quantities reflecting different time domains. 
The spin of gas reflects the sAM of recently accreted cold gas, while the spin of the dark matter halo is an integration of the full assembly history. 
Therefore variations in the incoming streams from the cosmic web affect $\lamgas$ more and $\lamh$ less. 
This is in line with our finding that $\lamgal$ and $\lamh$ are particularly uncorrelated at high $z$, when both gas accretion and depletion (star formation) are much faster.

\smallskip
Second, gas ejection from the galaxy center due to stellar feedback tends to remove low-spin gas, mixes the gas in the hot halo, and recycles the gas if it cools (e.g., \citealt{defelippis17}). 
Gas outflows can also occur from the disk outskirts, which removes high-spin gas that returns with low spin, thus lowering the overall spin (DeGraf et al. in prep.).
Hence, stronger feedback generally means more AM exchange between the inner and outer halo.
However, stronger feedback also means less clumpiness, so the processes that facilitate AM transfer such as dynamical friction, ram pressure, and torques generated by the perturbed disk under violent disk instabitliy \citep{dsc09} may be less efficient.
We note that the NIHAO simulations, which have stronger stellar feedback than VELA, show better alignment between spin vectors, and marginally stronger $\lamgal$-$\lamh$ correlation. 
Therefore, it seems that the net effect of a strong feedback is working {\it for} a better correlation or alignment. 
However, this interpretation is hindered by the fact that NIHAO has poorer resolution, which also reduces the clumpiness of the galaxies \citep{buck17}.

\smallskip
Third, torques from the stellar disk on the inspiraling gas ring can spin down the galaxy, not affecting $\lamh$.
Balancing this effect, compaction gives rise to a central bulge, and thus less torque on the inspiraling gas and less angular momentum loss.

\smallskip
To conclude, the evolution of $\lamgal$ is the net effect of many coupled processes, and the null correlation between $\lamgal$ and $\lamh$ is not very surprising.




\section{Galaxy size predictor revisited}
\label{sec:size}

\smallskip
What we have learnt so far poses a challenge to the classic galaxy-size predictor 
\be
\label{eq:Re}
\Re \simeq \fj \lamh \Rv.
\ee 
We showed that the proportionality factor $\fj$ ($\equiv\lamgal/\lamh$) has very large scatter, and can vary on short time scales due to compaction, mergers, and other processes. 
The factor $\fj$ depends systematically on $\lamh$: when parametrized as $\fj\propto\lamh^{-b}$, the simulations show that $b\approx-1$ at high $z$, suggesting that $\lamh$ is not a good predictor of galaxy size, at least at $z\ga1$.

\begin{figure*}
\includegraphics[width=0.75\textwidth]{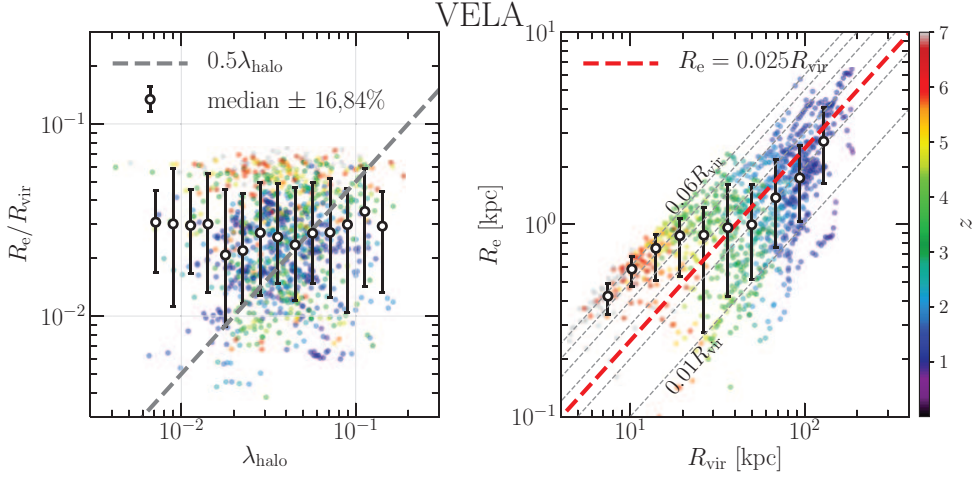} 
\caption{
Galaxy size (3D half-stellar mass radius) to host halo virial radius ratio versus halo spin ({\it left}), and 
galaxy size versus host halo virial radius ({\it Right:}), colorcoded by redshift, in the VELA simulations.  
Circles with errorbars indicate the median and 16 and 84 percentiles.
Dashed line in the left-hand panel represents a reference line, $\Re=0.5\lamh\Rv$. 
\citet{somerville17} showed that this relation, when combined with the stellar to halo mass relation from abundance matching, reproduces the observed $\Re$-$\Mstar$ relations across redshift up to $z\sim3$. 
Dotted lines in the right-hand panel are loci of $\Re = A \Rv$, with $A =$ 0.01,0.02,...,0.06. 
The overall best-fit $A$ is 0.025 (red, dashed line), while obviously $A$ increases with increasing redshift, in qualitative agreement with observation. 
}
\label{fig:size_VELA} 
\end{figure*}

\begin{figure*}
\includegraphics[width=0.75\textwidth]{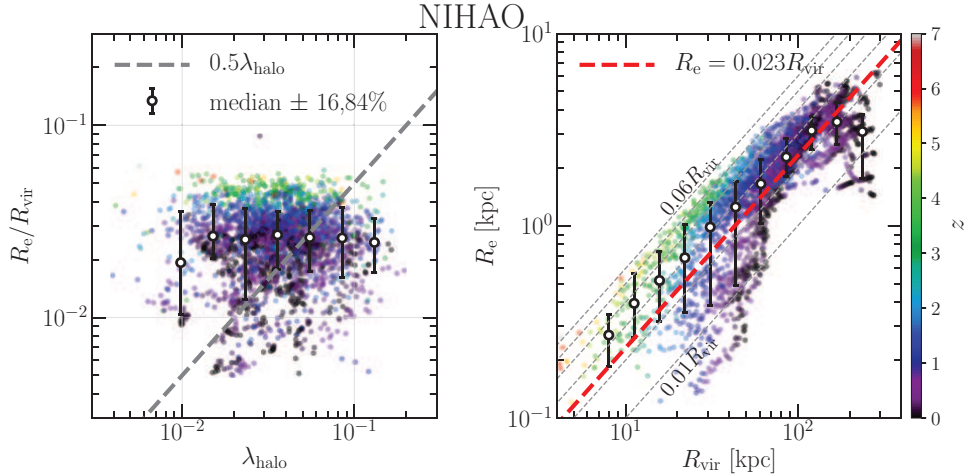} 
\caption{
The same as \fig{size_VELA}, but for the NIHAO simulation. 
Overall, $\Re = A \Rv$ with $A=0.023$, similar to that of VELA.  
The redshift dependence of $A$ is in qualitative agreement with VELA.
}
\label{fig:size_NIHAO} 
\end{figure*}

\smallskip
Here we check the validity of \equ{Re} with the VELA and NIHAO simulations.
The left-hand panels of \figs{size_VELA}{size_NIHAO} show $\Re/\Rv$ versus $\lamh$ for the VELA and NIHAO galaxies across all snapshots at $z<7$.
Note that in \fig{size_NIHAO} and for the rest of the paper, we have included the full NIHAO sample of $\sim100$ galaxies whose $\Mv$ range from $10^{9.5}\Msun$ to $10^{12.5}\Msun$. 
The full sample, with respect to the MW-sized subsample, shows the same trends regarding $\Re$ versus $\Rv$, and helps with the statistics.

\smallskip
There is almost no correlation between $\Re/\Rv$ and $\lamh$, independent of redshift. 
Interestingly, the right-hand panels of \figs{size_VELA}{size_NIHAO} show that both simulation suites still nicely reproduce the observed relation $\Re \approx A \Rv$, with the overall best-fit $A\approx0.024$.
In addition, both suites show a clear redshift trend: the proportionality factor $A$ increases from $\simeq0.02$ at $z\la1$ to 0.03-0.04 at $z\sim3$, and to approximately 0.05 at $z\ga5$.
This redshift trend qualitatively agrees with what \citet{somerville17} found using the GAMA and CANDELS surveys and abundance matching, but is a bit too strong. 


\smallskip
That is, in the simulations, the relation $\Re \approx A\Rv$ naturally arises, although $A$ is not a function of $\lamh$.
The interesting questions become: 
{\it what determines the value of $A$? Are there any secondary halo properties (secondary to halo mass or $\Rv$) that can capture the scatter in $A$? What gives rise to the redshift dependence? }
Our goal here is to generalize from the simulations an empirical recipe for predicting galaxy size using solely halo properties that can be useful for semi-analytic or semi-empirical models.

\subsection{a new empirical galaxy size predictor}

\smallskip
It turns out that, at fixed halo mass, smaller galaxies tend to live in more concentrated haloes, where the halo concentration parameter is defined as $c\equiv\Rv/\Rs$, with $\Rs$ the scale radius of the best-fit \cite{nfw97} profile. 
We measure the concentration parameter by fitting an NFW circular velocity profile to the circular velocity profile $\Vc(r)$ of the dark matter component of the simulated galaxy, as detailed in Appendix \ref{sec:concentration}.
Assuming for simplicity a power-law dependence on $c$, we find that galaxy size scales with halo radius and concentration as
\be
\label{eq:Re_new}
\Re = A^\prime c^\gamma  \Rv,
\ee 
with $\gamma\approx-0.7$ in both simulation suites.
As such, \equ{Re_new} is a tighter relation than $\Re = A\Rv$, and the factor $A^\prime$ is almost independent of redshift or mass.

\smallskip
We note that the role of the concentration dependence is two-fold.
First, at fixed halo mass and redshift, the size of {\it individual} galaxies anti-correlates with halo concentration.
The anti-correlation is well approximated by $c^{-0.7}$.
We illustrate this point in Appendix \ref{sec:cDependence}.
Second, there is a redshift dependence associated with $c^{-0.7}$, which captures the evolution of the average $\Re$-to-$\Rv$ ratio. 
It is well established that halo concentration is a function of halo mass and redshift. 
Using $N$-body simulations of the Planck cosmology, \citet{dutton14} provide an empirical concentration-mass-reshift relation, given by
\be \label{eq:concentration}
\log \langle c\rangle = a + b\log(\Mv/10^{12}\Msunh)
\ee
where $a=0.537+0.488\exp(-0.718z^{1.08})$ and $b=-0.097+0.024z$.
For $\Mv\sim 10^{12}\Msun$, this relation is well approximated by $\langle c\rangle \propto (1+z)^{-0.75}$ up to $z\sim3$.
Therefore, with the factor $c^{-0.7}$, \equ{Re_new} indicates that $A\propto(1+z)^{0.5}$, as found in both VELA and NIHAO.
This is illustrated in more detail in Appendix \ref{sec:comparison}.

\smallskip
\fig{size_concentration_VELA} and \fig{size_concentration_NIHAO} show $\Re$ versus $\Rv$ in the left-hand panels, and $\Re$ versus the concentration-corrected halo radius, $(c/10)^{-0.7}\Rv$, in the right-hand panels, for VELA and NIHAO respectively. 
Clearly the concentration scaling leads to a tighter and more universal relation for both suites. 

\smallskip
From the perspective of semi-analytic models, which build upon DMO simulations, we also check the validity of \equ{Re_new} using concentrations and virial radii measured from the matching DMO snapshots of the NIHAO simulations. 
This is shown in \fig{size_concentration_NIHAO_DMO}.
Comparing \fig{size_concentration_NIHAO} and \fig{size_concentration_NIHAO_DMO}, we can see that the same recipe holds, although the best-fit $A^\prime$ is somewhat different, reflecting the average halo response to baryonic physics.
We conclude that galaxy half mass radius {\it in the simulations} can be empirically modeled as $A^\prime (c/10)^{-0.7}\Rv$, with $A^\prime$ of the order of 0.02 and slightly dependent on the details of baryonic physics.

\smallskip
Intuitively, the dependence of galaxy size on halo concentration can be rationalized as follows. 
When considering a fixed halo mass, the concentration measures the depth of the gravitational potential well.
For an NFW profile, $\Phi_0 = -\Vv^2 c/f(c)$, where $\Phi_0$ is the gravitational potential at the center of the halo, $\Vv$ is the virial velocity, and $f(x)=\ln(1+x) - x/(1+x)$. 
In this regard, the success of \equ{Re_new} simply indicates that galaxy size contains information about the depth of the host potential, such that smaller galaxies live in deeper potential wells, 
possibly due to the fact that these haloes tend to form earlier (e.g., \citealt{wechsler02,zhao09}). 
While the above reasoning is suggestive, the cause of the concentration dependence of $\Re/\Rv$ is posed here as a theoretical challenge, especially the origin of the slope $\gamma \simeq -0.7$ in \equ{Re_new}. 

\begin{figure*}
\includegraphics[width=0.75\textwidth]{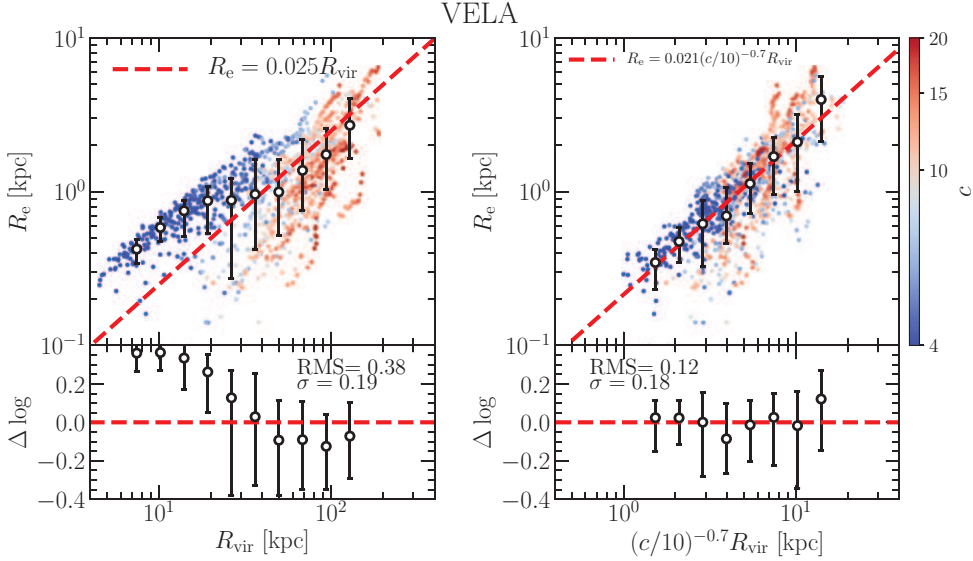} 
\caption{
{\it Left}: galaxy size (3D half-stellar mass radius) $\Re$ versus halo virial radius $\Rv$, for the VELA simulations throughout redshifts ($z=0.8-7$), colorcoded by halo concentration.
The red, dashed line is the best-fit relation of the functional form of $\Re=A\Rv$, as indicated.
{\it Right}: galaxy size $\Re$ versus the concentration-corrected halo radius, $(c/10)^{-0.7}\Rv$. 
The red, dashed line is the best-fit relation of the form of $\Re=A^\prime c^{\gamma}\Rv$, with $\gamma=-0.7$ fixed.
Circles with error bars indicate the median and the 16th and 84th percentiles. 
The {\it bottom} panels show the residual with respect to the best-fit model.
The quoted numbers are: the root-mean-square of the medians with respect to the best-fit models, and the average 1$\sigma$ scatter of $\Re$ in the bins of halo radius.
The concentration scaling makes the relation between galaxy size and halo radius tighter and more universal. 
}
\label{fig:size_concentration_VELA} 
\end{figure*}
\begin{figure*}
\includegraphics[width=0.75\textwidth]{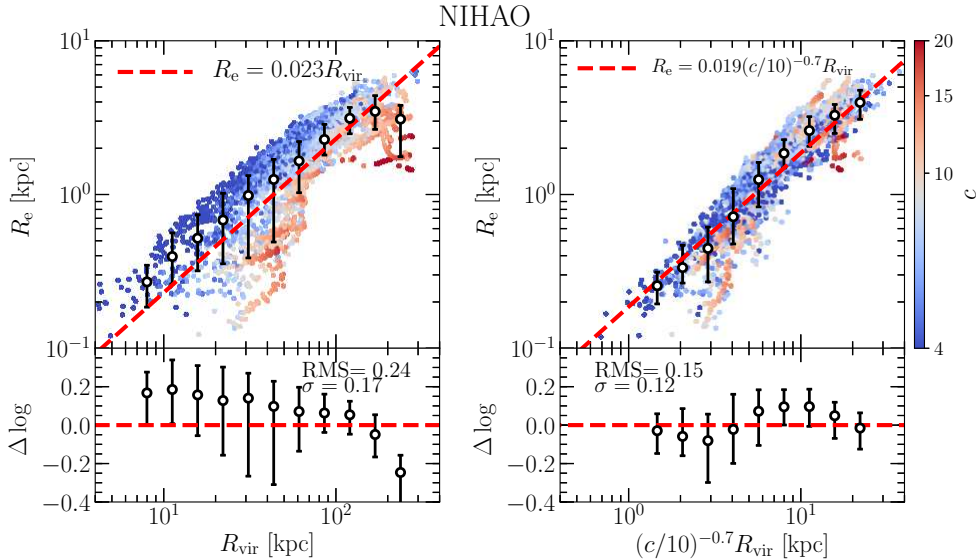} 
\caption{
The same as \fig{size_VELA}, but for the NIHAO simulations ($z=0-7$). 
The same concentration-scaling, $\Re\propto c^{-0.7}\Rv$, works equally well for NIHAO, with a similar zero-point as found in VELA. 
}
\label{fig:size_concentration_NIHAO} 
\end{figure*}
\begin{figure*}
\includegraphics[width=0.75\textwidth]{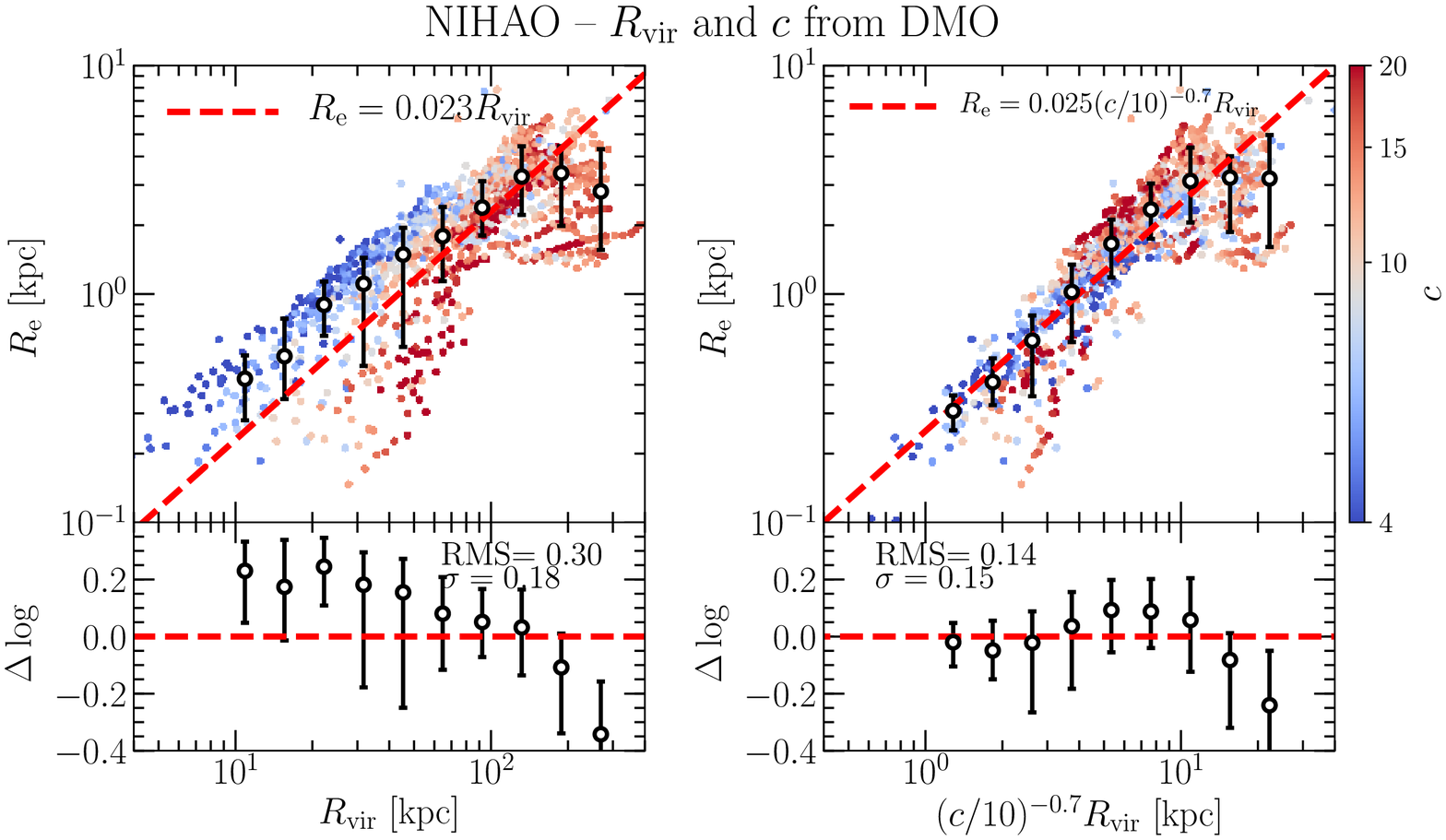} 
\caption{
Similar to \fig{size_concentration_NIHAO}, but with the halo properties measured in the matching dark-matter-only simulations. 
The data points are sparser than in \fig{size_concentration_NIHAO} because for every four hydro output snapshots there is only one dark-matter-only output.
The same empirical relation $\Re\propto c^{-0.7}\Rv$ holds, although the scatter is somewhat larger.  
The difference between the zero point here (0.025) and in \fig{size_concentration_NIHAO} (0.019) reflects the average halo response to baryonic processes in the NIHAO simulations. 
}
\label{fig:size_concentration_NIHAO_DMO} 
\end{figure*}
%

\section{Discussion}
\label{sec:discussion}

\subsection{Comparison with previous studies}
\label{sec:discussion1}

\smallskip
In this section, we compare our results regarding the correlations of spin amplitudes and spin vectors with those reported in the literature. 

\smallskip
\cite{teklu15} use the Magneticum Pathfinder simulations to study the connection between the kinematic morphology of galaxies and their specific baryonic AM and host halo spin.
Although their primary focus is the morphology dependence, Fig.11 therein actually shows that the sAM of the halo $\jh$ and the sAM of the cold gas $\jgas$ are barely correlated at $z>1$, for either disks or spheroids.
This is in agreement with what we find in VELA and NIHAO. 

\smallskip
In addition, \cite{teklu15} show that at lower redshifts there is a weak correlation coming into place, consistent with what NIHAO implies.
We have verified using the Illustris simulations \citep{genel14} that at $z\ga1$ the spin of baryons and the spin of the host halo are barely correlated, and that at lower-$z$ there is a weak correlation developing primarily between $\lams$ and $\lamh$.
The low-$z$ behavior is also confirmed by \cite{rg17}, who report a correlation in the Illustris simulation between the degree of rotation-support of stars and host halo spin for $z=0$ galaxies with $\Mstar<10^{11}\Msun$.

\smallskip
We note that both the Magneticum Pathfinder simulations and the Illustris simulations have taken AGNs into account, and exhibit similar results compared to VELA and NIHAO. 
Therefore, AGNs seem to have rather weak effect regarding the $\lamgal$-$\lamh$ correlation. 


\smallskip
We find the alignment of galaxy spin and halo spin to be marginally weaker at later times, and the alignment of gas spin and stellar spin to become slightly better at later times. 
The same qualitative trends are found by \citet{zs17} in the Illustris simulation, where the spins of dark matter, gas and stars are all measured within the whole virial radius.

\smallskip
The median angle between the spin vectors of the cold gas and the host halo, $\langle \theta_{\rm gas, halo}\rangle$, is 43-45$^\circ$ in the NIHAO simulations, weakly dependent on redshift. 
This is in good agreement with most of the reported values in the literature: \cite{hahn10} found 49$^\circ$ at $z=0$; \cite{teklu15} found 45-49$^\circ$ at $z=0.1$, depending weakly on galaxy morphology. 
\cite{sharma12} reported a significantly smaller value of 30$^\circ$.
The median $\langle \theta_{\rm gas, halo}\rangle$ is 47-54$^\circ$ for VELA galaxies with $\Mv>10^{11.4}\Msun$, and is 51-55$^\circ$ for $\Mv<10^{11.4}\Msun$, depending on redshfit.
These results are on the high side of the literature values.

\smallskip
The median angle between the spin vectors of the stars and the host halo $\langle \theta_{\rm stars, halo}\rangle = 46$-$50^\circ$ in the NIHAO simulations, also in good agreement with previous studies.
\cite{croft09} found 44$^\circ$ at $z=1$; \cite{hahn10} measured 49$^\circ$ at $z=0$; \cite{teklu15} found 46-57$^\circ$ at $z=0.1$;
\cite{bett10} reported a significantly smaller value of 34$^\circ$ at $z=0$; in comparison, at $z<0.8$, $\langle \theta_{\rm stars, halo} \rangle=49^\circ$ in NIHAO.
VELA results are again on the high side, with $\langle \theta_{\rm stars, halo} \rangle=61^\circ$ at $z=0.8-2$. 

\smallskip
By tracing the Lagrangian volumes of the inner 10\% of the virial radii at $z=0$, \cite{zavala16} found that the angular momentum loss of baryons tightly correlates with that of the dark matter, since the turn-around time of the dark matter.
This is in line with our finding that the spins of the inner halo and the galaxy are correlated. 

\smallskip
\cite{hahn10} and \cite{teklu15} found the median angle between the spin vectors of the baryonic components to be $\langle\theta_{\rm gas,stars}\rangle \approx6-8^\circ$ for late-type galaxies, with very weak redshift dependence in the range $z=$0-2. 
This is bracketed by the NIHAO result of 5$^\circ$ at $z<2$ and the VELA result of 13$^\circ$ at $0.8<z<2$. 

\smallskip
Starkenburg et al. (2018, in prep) study the counter-rotating galaxies in the Illustris simulation in detail. 
They find the counter-rotating fraction to be very low. 
The fraction of galaxies with the angle between the spin vectors of the gas and stars larger than 120$^\circ$ is 0.43\%, independent of halo mass for $\Mv=10^{11.4-12.2}\Msun$ and decreasing at higher masses (private communication).
Note that the Illustris sample is generally more massive than the galaxies used in this study, so it is possible that counter-rotations are more common in low mass systems. 
In fact, for $\Mv>10^{11.4}\Msun$, no counter-rotation defined in the same way is detected in VELA, consistent with Starkenburg et al.

\subsection{Redshift dependence of $\Re$ in comparison with observations}
\label{sec:discussion2}

\smallskip
We note that, while \equ{Re_new} with $\gamma\simeq-0.7$ accurately describes the simulation results, the implied redshift dependence of $\Re/\Rv$ seems too strong compared to that inferred from halo abundance matching \citep{somerville17}. 
In particular, the concentration scaling $c^{-0.7}$ yields approximately $\Re/\Rv \propto (1+z)^{0.5}$, while the observationally inferred redshift trend is approximately $\propto (1+z)^{0.3}$.
This is illustrated in Appendix \ref{sec:comparison}.

\smallskip
The key observational benchmark for a galaxy size predictor is the $\Re$--$\Ms$ relation, which exhibits clear redshift evolution such that, at fixed $\Mstar$, galaxies are more compact at higher $z$ (e.g., \citealt{vanderwel14_MR}; \citealt{somerville17}).
In the context of empirical modeling of observations, the prediction of the $\Re$-$\Mstar$ relation requires the combination of the relation between the stellar mass and halo mass from abundance matching and the relation between galaxy size and halo radius from theory.
In practice, starting from the halo catalog of an $N$-body simulation, one converts $\Mv$ to $\Ms$ using the $\Ms$-$\Mv$ relation, and computes $\Re$ using $\Rv$ and halo structural parameters ($\lamh$, $c$) according to the size predictor.

\smallskip
As said, \equ{Re_new} exhibits a $z$-dependence that is too strong.
At $z\sim2$, the average galaxy size predicted by $\Re = 0.02(c/10)^{-0.7}\Rv$ is $\sim$50\% higher than the 3D-half mass radius observed in CANDELS deduced by \citet{somerville17}.

\smallskip
We opt not to interpret this tension too literally, for two reasons. 
First, the half-mass radius as deduced from the observed 2D radius may be biased as a function of redshift, and second, the $\Ms$--$\Mv$ relation may carry non-negligible uncertainties, as follows.

The deprojection from 2D to 3D may be strongly biased. The conversion of the projected half-light radius ($R_{\rm e,2D}$) to the 3D half-stellar-mass radius ($\Re$) involves two factors:
\be\label{eq:2Dto3D}
R_{\rm e,2D} = f_{\rm p} f_{\rm k} \Re,
\ee
where $f_{\rm p}$ corrects for projection and $f_{\rm k}$ accounts for the convertion from light-weighting to mass-weighting.
While $f_{\rm k}$($\sim1.2$) seems to be a weak function of mass and redshift \citep{dutton11, lange15}, $f_{\rm p}$ depends significantly on the structure and shape of galaxies and is very likely a strong function of redshift and mass. 
For spherically-symmetric spheroids or oblate systems, $f_{\rm p}$ is typically between 0.68 (de Vaucouleurs) and 1 (face on exponential disk), but for elongated (prolate) systems, $f_{\rm p}$ can easily be much smaller than unity. 
Qualitatively, this can be comprehended by considering a cigar-shaped galaxy projected along the major axis.
It is easy to show that ellipsoids with an intrinsic axis ratio of $b/a\sim0.4$, $f_{\rm p}$ has a value of $\sim0.5$ for exponential density profiles.
The fraction of prolate galaxies increases towards high redshifts and lower masses: at $z\sim1$, more than half of all the galaxies with $\Ms\sim10^{9}\Msun$ are prolate (\citealt{vanderwel14_MR}, Zhang et al. in prep). 
This is supported by simulations.
In fact, VELA galaxies typically have $b/a\sim0.4$ at $z=2$--2.5 \citep{ceverino15_shape,tomassetti16}.

\smallskip
\citet{somerville17}  practically applied $f_{\rm p}f_{\rm k}=1\times1.2 = 1.2$ for late-type galaxies and $f_{\rm p}f_{\rm k}=0.68 \times 1.15 = 0.78$ for early-type galaxies. 
That is, for the relevant mass range ($\Ms\la10^{10.5}\Msun$) where late-type galaxies dominate, $\Re$ is approximately always 20\% smaller than $R_{\rm e,2D}$, while for prolate systems, which dominate at high redshifts, it should be significantly larger than $R_{\rm e,2D}$.
Taking the effect of projecting elongated systems into account will increase the $\Re$ deduced from observations and potentially alleviate the tension. 

\smallskip
The $\Ms$-$\Mv$ relations may carry non-negligible systematic uncertainties. 
Almost all the abundance matching studies assume a \citet{chabrier03} initial mass function (IMF) as found for the Milky Way throughout redshifts.
However, the IMF may not be universal.
In fact, even in the local universe, there is no consensus whether the IMF is Milky-Way like or bottom heavy (e.g., \citealt{dutton12}).
If the IMF is closer to Salpeter (1955) at $z\sim2$, the stellar mass would be higher by $\sim$0.25dex with respect to the standard abundance matching result based on the Chabrier IMF.
This alone almost adequately accounts for the 50\% overprediction of the $\Re$-$\Ms$ relation.

\smallskip
If the goal, despite these caveats, is to reproduce the observational estimates as they are, using the standard abundance matching result, one can simply introduce an extra redshift-dependence (in addition to and in the opposite direction to that associated with $c(z)^{-0.7}$) by generalizing \equ{Re_new} to 
\be \label{eq:Re_new2}
\Re = A^\prime (1+z)^{\beta} c^{\gamma} \Rv.
\ee
We find that $\beta\approx-0.2$ serves as an adequate correction for $\gamma=-0.7$.


We clarify again that \equ{Re_new2} is a deviation from \equ{Re_new} that accurately describe the redshift dependence in the simulations.
\Equ{Re_new2} should be adopted if the tension between the simulated sizes and the sizes deduced from observations at high redshift is confirmed to be valid. 

\subsection{Other comments}
\label{sec:discussion3}

\smallskip
We also note that, although the concentration parameter has been used in conventional galaxy size predictors of the form of \equ{Re}, 
its role was rather limited.
In recipes of the sort of \citet{mmw98}, concentration cames in via a factor of order unity that describes the adiabatic contraction of the halo due to the gravity from the galactic disk.
Here, with \equ{Re_new2} and $\gamma\approx-0.7$, the $c$-dependence is more pronounced.

\smallskip
The concentration dependence may also {\it partially} explain the morphology dependence of the $\Re$-$\Rv$ relation.
In the context of conditional abundance matching, the specific star formation rate (color) of a galaxy contains information of the host halo formation time, in the sense that
older galaxies dwell in haloes that formed earlier at a given mass \citep{hearin13}. 
Galaxy color correlates with morphology, and concentration reflects halo formation time.
That is, \equ{Re_new2} hints that quiescent galaxies are more compact at a given halo mass (virial radius) than star-forming ones. 

\smallskip
The most prominent difference between VELA and NIHAO in \figs{size_concentration_VELA}{size_concentration_NIHAO} is that the scatter is larger in VELA for both the $\Re-\Rv$ relation and the $\Re$-$c^{-0.7}\Rv$ relation. 
This is likely because the NIHAO sample consists of almost exclusively large, late-type galaxies, while the VELA suite covers a wider range of morphologies.
The fact that the $\Re$--$c^{-0.7}\Rv$ relation still exhibits significant scatter, hints that there might be residual dependence of $\Re$ on morphological type.
It remains an interesting open question for future studies, whether additional halo properties can help tighten the relation between galaxy size and halo virial radius further.

\section{Conclusion}
\label{sec:conclusion}

\smallskip
In this paper, we use two suites of cosmological hydrodynamical simulations to study the correlation between galaxy spin and its host halo spin, and examine the relation between galaxy effective radius and halo virial radius.
The two suites differ significantly in numerical resolution and in the strength of stellar feedback, yet show similar results regarding the correlation of spins and the galaxy size - halo size relation, from which we draw the following conclusions.

\smallskip
\noindent
(i) The distribution of galaxy spin follow similar log-normal shapes to that of their host haloes. 
The median values differ for different components: the spin of stars within $0.1\Rv$ has the value of $\sim0.005$--$0.007$, while the spin of cold gas within $0.1\Rv$ is $\sim0.02$--$0.03$, slightly lower than but of the same order of the dark matter halo spin ($\sim0.037$), in qualitatively agreement with what is inferred from the $H_\alpha$ kinematics of massive star forming galaxies at high-$z$ \citep{burkert16}.

\smallskip
\noindent
(ii) The similarity of the spin distributions does not translate to a correlation between the spins of each given galaxy and the spins of its host halo.
Both simulation suites show that galaxy spin $\lamgal$ and host halo spin $\lamh$ are barely correlated, especially at $z\ga1$. 
This null correlation is qualitatively the same, if the spin of the galaxy is measured for the cold gas or the stars separately. 
There seems to be a weak correlation between $\lamgal$ and  $\lamh$ at $z\la1$.
Given that the specific angular momentum of cold gas and dark matter are correlated at the accretion into the halo virial radius, this indicates that the gas angular momentum is not conserved during the gas inflow into the galaxy and its evolution within the galaxy. 
The spin of the inner part of the dark matter halo shows a correlation with that of the galaxy, but this is a consequence of baryonic effect on the dark matter halo, such that the inner halo spin from $N$-body simulations cannot serve as a proxy for the galaxy spin.

\smallskip
\noindent
(iii) The angular momentum retention factor, $\fj$ ($\equiv \lamgal/\lamh$), has a value of $\sim0.5$ on average, with large, stochastic variations from galaxy to galaxy and from one time to anther, and it anti-correlates with $\lamh$. 
Wet compaction or mergers could potentially give rise to the anti-correlation between $\fj$ and $\lamh$. 
Low-$\lamh$ galaxies tend to develop a wet compaction \citep{db14}, which causes the initially low-spin system to end up with higher $\lamgas$ by depleting the low-angular momentum gas in the compact star forming nucleus phase, and acquiring higher-angular momentum gas that settles in an extended ring.
The merger of massive satellites causes $\lamh$ and $\lamgal$ to rise and fall {\it in turn}, over a time scale of several halo dynamical times.
Based on the picture of angular momentum gain and loss described in \cite{danovich15}, we also speculated on other possible mechanisms, which do not necessarily cause an anti-correlation between $\fj$ and $\lamh$, but smear out the $\lamgal$-$\lamh$ correlation. 

\smallskip
\noindent
(iv) Contrary to the uncorrelated spin amplitudes, the spin orientations of a galaxy and its host halo are correlated.
Overall, half of the cases have $\cos\theta\ga 0.6$, where $\theta$ is the angle between the galaxy spin vector and halo spin vector.
At a given halo mass bin, the alignment becomes marginally weaker at later times.
This suggests that the mechanisms that smear out the correlation of the amplitudes of spin do not totally randomize the alignment.
The spin alignment is consistent with the finding that the inflow is predominantly in a preferred plane \citep{danovich12}, such that the torques exerted are preferentially along the spin vector, thus affecting its amplitude but not its direction.
 
\smallskip
\noindent
(v) The NIHAO simulations, which have stronger stellar feedback than the VELA simulations, show a marginally stronger correlation between the spin of the galaxy and of its host halo, and a slightly better alignment between the spin vectors. 
This may suggest that stronger feedback operates to the advantage of a better correlation, via stronger angular momentum exchange between the inner part and outskirts of a galaxy. 
We caution though that this interpretation is hindered by the fact that NIHAO has poorer resolution, which, as with strong feedback, also reduces the clumpiness of galaxies and thus is disadvantageous to angular momentum exchange.

\smallskip
\noindent
(vi) The spins of the cold gas and the stars in a galaxy are correlated, with $\mathcal{R}\sim0.6-0.8$, strengthening at later times. 
The spin vectors of the baryonic components are well aligned, with about half of the cases having $\cos\theta_{\rm gas, star}\ga0.97$.  
The alignment is better for more massive systems. 
A non-negligible fraction of the cases have counter-rotating gas and stars, especially in lower-mass systems.

\smallskip
\noindent
(vii) We find that the halo spin parameter is not significantly correlated with galaxy size in both simulation suites, challenging the conventional, semi-analytic galaxy size estimator: $\Re\simeq 0.5 \lamh\Rv$. 
Nevertheless, both VELA and NIHAO reproduce the empirical relation derived from abundance matching, $\Re\simeq A\Rv$, with the proportionality factor $A\simeq0.02$ at low-$z$ and increasing towards high-$z$.
We find that in the simulations, galaxy size can be well described by the relation 
\be\label{eq:Re_new_conclusion}
\Re = 0.02 (c/10)^{-0.7} \Rv
\ee 
where $c$ is the halo concentration. 
The concentration dependence serves two purposes.
First, at fixed halo mass and redshift, the size of individual galaxies anti-correlates with halo concentration as $c^{-0.7}$.
Second, there is a redshift dependence associated with the $c^{-0.7}$ factor, which comes in via the concentration-mass-redshift relation and accurately captures the redshift evolution of $A$ in the simulations.
This redshift trend, however, is too strong compared to that inferred from observation using abundance matching and a simplified deprojection of sizes, and thus seems to cause over-prediction of galaxy size at high redshift at given $\Mstar$.  
Although there might be some caveats that could potentially change the results deduced from observations, if reproducing the observed $\Re$-$\Mstar$ relations across redshifts is the primary concern, one can apply an extra redshift dependence to the size predictor, making it $\Re = 0.02 (1+z)^{-0.2} (c/10)^{-0.7} \Rv$.
We clarify again that this is a deviation from \equ{Re_new_conclusion} that accurately describe the redshift dependence in the simulations, and should be adopted if the tension between the simulated sizes and the sizes deduced from observations at high redshift is confirmed to be valid. 
This will imply that the simulated radii are inaccurate at high redshift. 
Alternatively, one can assume that the tension is an artifact of the process of analyzing the simulations, that the simulations are reliable, and adopt \equ{Re_new_conclusion}. 
The empirical relation can be tested with future observations where concentration and virial radius are measured from gravitational lensing.
It remains an open question how the potential well of the host halo regulates the size of the galaxy to give rise to the specific $c^{-0.7}$ scaling. 

\section*{Acknowledgments}

\smallskip
We acknowledge stimulating discussions with Andreas Burkert and Reinhard Genzel.
This work was partly supported by the grants ISF 124/12, I-CORE Program of the
PBC/ISF 1829/12, BSF 2014-273, PICS 2015-18, and NSF AST-1405962.
FJ is supported by the Planning and Budgeting Committee (PBC) fellowship of the Council for Higher Education in Israel. 
JP is support by the grant HST-AR-14578.001-A.
The VELA simulations were performed at the National Energy Research Scientific Computing Center (NERSC) at Lawrence Berkeley National Laboratory, and at NASA Advanced Supercomputing (NAS) at NASA Ames Research Center. 
DC is supported by the ERC Advanced Grant, STARLIGHT: Formation of the First Stars (project number 339177). 
The NIHAO simulations were performed on the High Performance Computing resources at New York University Abu Dhabi; on the THEO cluster of the Max-Planck-Institut f${\ddot{u}}$r Astronomie and on the HYDRA clusters at the Rechenzentrum in Garching.

\bibliographystyle{mn2e}
\bibliography{bn}

\appendix

\section{Halo concentration measurement}
\label{sec:concentration}

\smallskip
We measure the halo concentration parameter by fitting the circular velocity profile $\Vc(r)=\sqrt{GM_{\rm dm}(<r)/r}$ of the simulated galaxies with an NFW profile, where $M_{\rm dm}(<r)$ is the dark matter mass within radius $r$. 
The circular velocity profile is measured in 30 spherical shells with equal thickness in logarithmic scale between 0.01-1$\Rv$.
In practice, we minimize the following figure of merit:
\be\label{eq:FitConcentration}
\rms^2 = \frac{1}{N} \sum_{i=1}^{N=30}\left[\frac{V_{\rm c,NFW}(x_i|\Mv,c)-\Vc(x_i)}{\Vc(x_i)}\right]^2
\ee
where 
\be\label{eq:FitConcentration}
V_{\rm c,NFW}(x|\Mv,c) = \sqrt{ G \frac{\Mv}{r} \frac{f(cx)}{f(c)} },
\ee
with $f(y)=\ln(1+y)-y/(1+y)$ and $x=r/\Rv$.

\smallskip
Note that for \fig{size_concentration_VELA} and \fig{size_concentration_NIHAO}, we have excluded the snapshots with $\rms >0.07$, corresponding to the worst 20\% fits, where the NFW profile is not a good description of the dark matter halo density profile. 
Usually the excluded cases are associated with major mergers.

\section{Concentration dependence at fixed halo mass and redshift}
\label{sec:cDependence}

\begin{figure}
\includegraphics[width=0.5\textwidth]{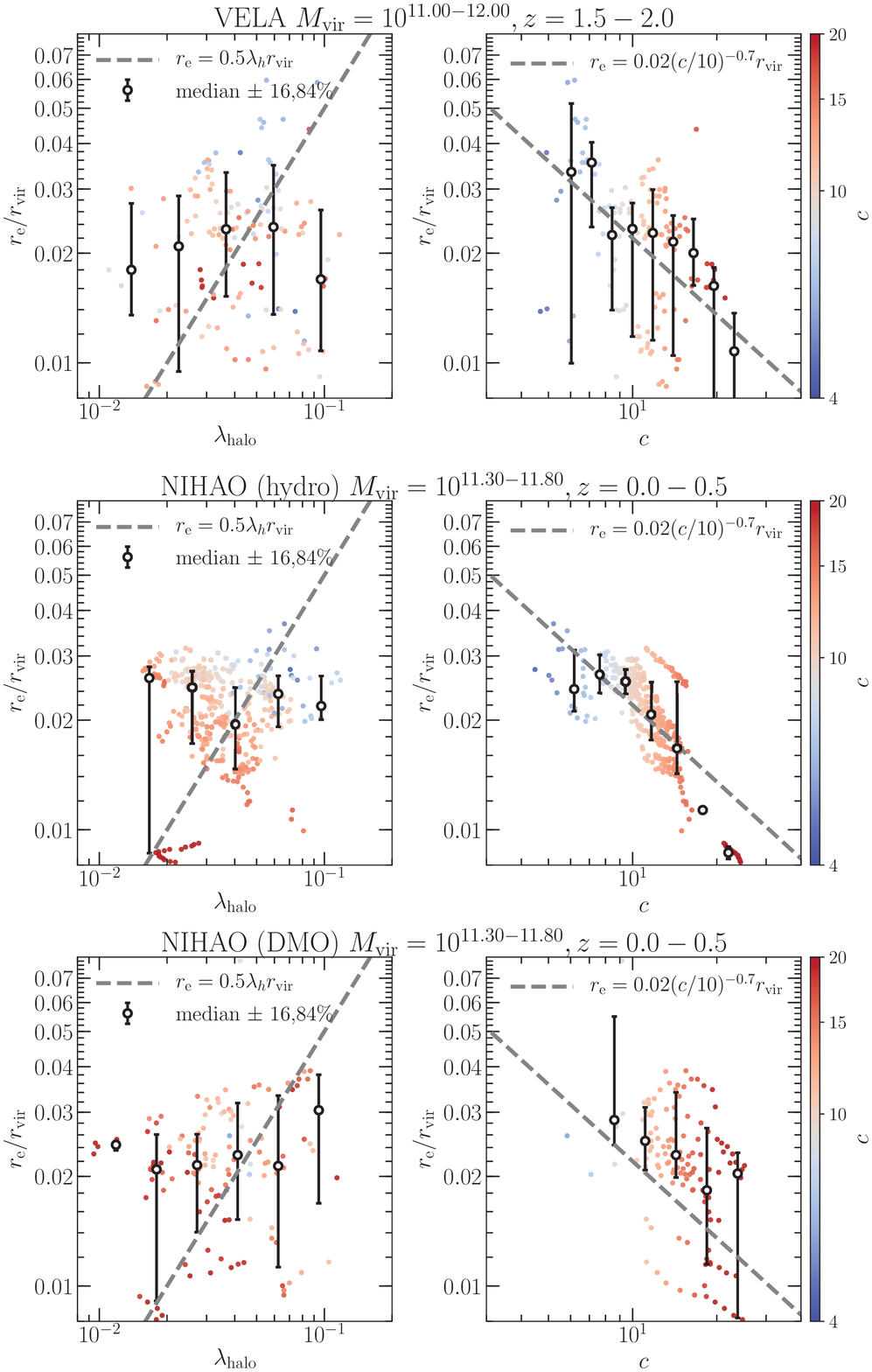} 
\caption{
The ratio $\Re/\Rv$ as a function of spin $\lamh$ ({\it left}) and concentration $c$ ({\it right}), in a narrow range of halo mass and redshift.
{\it Top} : VELA galaxies with $\Mvir=10^{11-12}\Msun$ at $z=1.5-2$.
{\it Middle} : NIHAO galaxies with $\Mvir=10^{11.3-11.8}\Msun$ at $z=0-0.5$.
{\it Bottom} : NIHAO galaxies with $\Mvir=10^{11.3-11.8}\Msun$ at $z=0-0.5$, with galaxy size measured from the fiducial hydro simulation, and halo properties measured from the matching dark-matter-only simulations.
The dashed reference lines represent $\Re=0.5\lamh \Rv$ and $\Re = 0.02(c/10)^{-0.7}\Rv$, as indicated.
The three rows all show: In a narrow range of redshift and halo mass, the ratio $\Re/\Rv$ is clearly dependent on halo concentration $c$, well described by $c^{-0.7}$ and almost independent of halo spin $\lamh$.
}
\label{fig:cDependence_FixedMassRedshift} 
\end{figure}

\smallskip
Here we illustrate that, at fixed halo mass and redshift, galaxy size anti-correlates with halo concentration in our simulations.

\smallskip
\fig{cDependence_FixedMassRedshift} shows the ratio $\Re/\Rv$ versus $\lamh$ and $c$, for simulated galaxies in a small range of halo mass and redshift. 
For individual galaxies (snapshots), the ratio $\Re/\Rv$ depends on halo concentration $c$, the trend of which is well described by $c^{-0.7}$; and does not depend on $\lamh$, in both VELA and NIHAO.
The qualitative trend is valid no matter the halo properties ($c$, $\lamh$, and $\Rv$) are measured from the matching dark-matter-only runs of the NIHAO suite or in the fiducial hydro simulations. 
\footnote{The data points are sparser in the DMO panels than in the hydro panels, because the output timesteps are thicker: for every 4 hydro outputs, there is one DMO output.}
Further ensuring that the $c$-dependence is not a mass trend or redshift trend, we have verified that in narrow ranges of $\Mv$ and $c$, the ratio $\Re/\Rv$ does not depend on $z$, and that in narrow ranges of $z$ and $c$, the ratio $\Re/\Rv$ does not depend on $\Mv$.

\section{Comparison of the median size-mass relations}
\label{sec:comparison}

\begin{figure}
\includegraphics[width=0.48\textwidth]{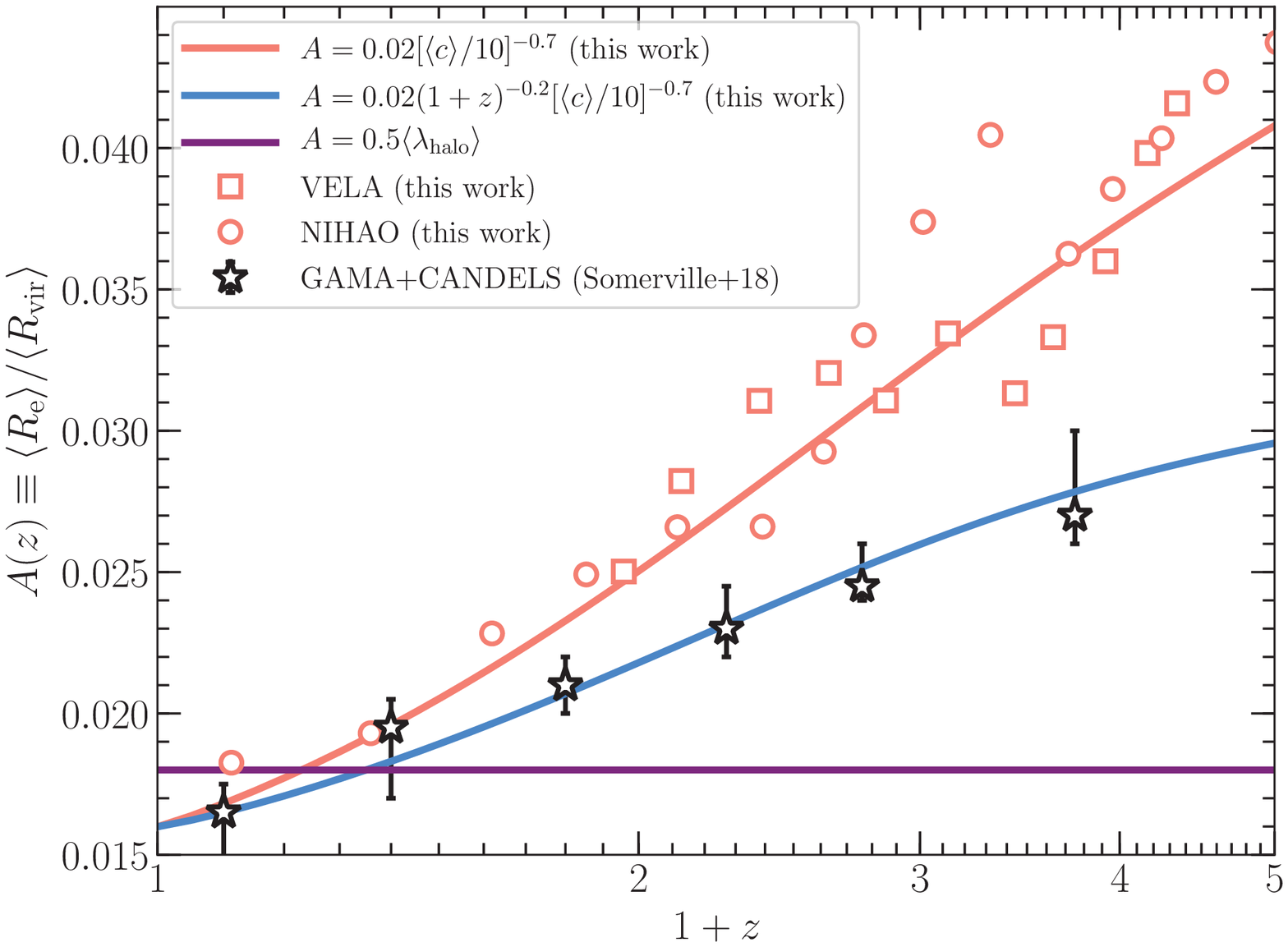} 
\caption{The zero-point of the median galaxy size - halo radius relation, $A(z)\equiv\langle \Re \rangle$/$\langle \Rv\rangle$,  as a function of redshift $z$. 
The orange line represents the model proposed in this work, i.e., $A(z) = 0.02[\langle c\rangle(\Mv,z)/10]^{-0.7}$, where $\langle c\rangle$ is given by the median concentration-mass-redshift relation of \citet{dutton14}, and we adopt $\Mv=10^{11}\Msun$, a typical halo mass in our simulations, for illustration. 
The squares and circles represent the VELA and NIHAO results, respectively. 
Note that for the simulation results, we have scaled the overall normalization up and down slightly just to align them roughly to better reveal the redshift trend.
The black stars represent observations (abundance matching) adopted from \citet{somerville17}, for galaxies with $\Mstar < 10^{10.5}\Msun$, with the errorbars indicate the full uncertainty in $\langle \Re \rangle$/$\langle \Rv\rangle$. 
The blue line represents $A(z) = 0.02(1+z)^{-0.2}[\langle c\rangle(\Mv,z)/10]^{-0.7}$.
The purple line represents $A=0.5\langle \lamh\rangle$, with $\langle \lamh\rangle=0.036$, assuming that $\langle \lamh\rangle$ does not evolve with redshift.
Obviously, the factor $c^{-0.7}$ introduces a $z$-dependence that matches the simulation results well but is steeper than that inferred from abundance matching.
To enforce an agreement with the abundance matching result, an extra $z$-dependence of $(1+z)^{-0.2}$ is required. 
}
\label{fig:Az} 
\end{figure}

\smallskip
In the new galaxy size predictor as given by $\Re = 0.02(c/10)^{-0.7}\Rv$, there are two roles of the concentration dependence. 
Here we illustrate one of them, that the average $\Re$-$\Rv$ relation evolves with redshift, and the redshift dependence as found in the simulations is well described by $c^{-0.7}$ through the concentration-mass-redshift relation. 
In Appendix \ref{sec:cDependence}, we illustrate the other role that,  at fixed redshift and halo mass, the size of individual galaxies in the simulations anti-correlates with $c$, and is well described by $c^{-0.7}$.

\smallskip
\fig{Az} shows the ratio $A(z)$ between the median galaxy size $\langle \Re\rangle$ and the median halo virial radius $\langle \Rv\rangle$ as a function of redshift $z$.
In our simulations, $A(z)$ increases by more than a factor of 2 from $z=0$ to $z=3$.
Using the concentration-mass-redshift relation from \cite{dutton14}, we find that $A\propto c^{-0.7}$ nicely captures this behavior. 

\smallskip
Observationally, however, the $z$-dependence is weaker: applying halo abundance matching to galaxies from the GAMA survey and the CANDELS survey, \cite{somerville17} find that $A$ increases by about 60\% from $z=0.1$ to $z=2.75$.
This is illustrated in \fig{Az} with the black stars.
Obviously, the $c^{-0.7}$ factor that well describes the $z$-dependence of the simulations overpredicts the $z$-dependence of the observational (abundance matching) $\Re$-$\Rv$ relations.
Therefore, if the relation $\Re = 0.02 (c/10)^{-0.7} \Rv$ is used in a semi-analytic or semi-empirical model, the model will not pass a key benchmark test -- the galaxy size-mass relation.

\smallskip
That said, if a perfect reproduction of the $z$-trend is needed, one can introduce empirically an extra $z$-dependence in the size predictor:
\be\label{eq:cModel}
\Re = f(z)\left[\langle c\rangle(\Mv,z)/10\right]^{-0.7} \Rv.
\ee
We find that with $f(z)=0.02(1+z)^{-0.2}$, \equ{cModel} reproduces the observationally deduced results quite well.
This can also be seen in \fig{Az}.

\smallskip
We emphasize that the $c$-dependence is introduced not only to reproduce the average $z$-dependence, but also to capture the $c$-dependence of the size of individual galaxies at fixed halo mass and redshift, in our simulations (Appendix \ref{sec:cDependence}). 
The $f(z)$ factor needed may reflect intrinsic inconsistencies between the simulations and observations.


\label{lastpage}
\end{document}